\documentclass[twocolumn]{aastex701}

\usepackage{cancel}
\usepackage{subfigure}
\usepackage{xcolor}
\usepackage{graphicx}
\usepackage{multirow}
\usepackage{array}
\usepackage{soul}
\usepackage{longtable}
\usepackage{rotating}
\usepackage{calc}
\usepackage{caption}
\usepackage[section]{placeins}

\shorttitle{GULP: Galaxy UV Legacy Project. I}
\shortauthors{Sabbi et al.}
\graphicspath{{./}{figures/}}

\begin{document}

\title{Galaxy UV Legacy Project: Survey Description and First Insights Into NGC~4449 Recent History of Star Formation}

\author[0000-0003-2954-7643]{E.\, Sabbi}
\affiliation{Gemini Observatory/NSFs NOIRLab, 950 N. Cherry Ave., Tucson, AZ 85719, USA}
\email{elena.sabbi@noirlab.edu}

\author[0000-0001-8658-2723]{B.\, Meena}
\affiliation{Space Telescope Science Institute, 3700 San Martin Dr, Baltimore, MD 21218, USA}
\email{bmeena@stsci.edu}

\author[0000-0002-6091-7924]{P.\, Zeidler}
\affil{AURA for the European Space Agency (ESA), ESA Office, Space Telescope Science Institute, 3700 San Martin Drive, Baltimore, MD 21218, USA}
\email{zeidler@stsci.edu}

\author[0009-0008-4009-3391]{V.\, Bajaj}
\affiliation{Space Telescope Science Institute, 3700 San Martin Dr, Baltimore, MD 21218, USA}
\email{vbajaj@stsci.edu}

\author[0000-0002-5189-8004] {D.\, Calzetti}
\affiliation{Dept. of Astronomy, University of Massachusetts, Amherst, MA 01003, USA}
\email{calzetti@astro.umass.edu}

\author[0000-0002-1722-6343]{J.\, J.\, Eldridge}
\affiliation{Department of Physics, The University of Auckland, Private Bag 92019, Auckland, New Zealand}
\email{j.eldridge@auckland.ac.nz }

\author[0009-0003-6044-3989]{P.\, Facchini}
\affiliation{Astronomisches Rechen-Institut, Zentrum f\"ur Astronomie der Universit\"at Heidelberg, M\"onchhofstr. 12-14, D-69120 Heidelberg, Germany}
\email{pietro.facchini@stud.uni-heidelberg.de }

\author[0000-0002-1000-6081]{S.\, Linden}
\affiliation{University of Arizona, Tucson, AZ 85719, USA}
\email{seanlinden@arizona.edu}

\author[0000-0001-6000-6920]{P.\, A.\, Crowther}
\affiliation{Astrophysics Research Cluster, School of Mathematical and Physical Sciences, University of Sheffield, Hounsfield Road, Sheffield, S3 7RH, United Kingdom}
\email{paul.crowther@sheffield.ac.uk}

\author[0000-0002-8192-8091]{A.\, Adamo} 
\affiliation{Department of Astronomy, Oskar Klein Centre, Stockholm University, AlbaNova University Centre, SE-106 91 Stockholm, Sweden}
\email{angela.adamo@astro.su.se}

\author[0000-0001-7746-5461]{L.\ Bianchi}
\affiliation{Department of Physics \& Astronomy, The Johns Hopkins University, 3400 N. Charles St., Baltimore, MD 21218, USA}
\email{bianchi@jhu.edu}

\author[0000-0001-6291-6813]{M.\, Cignoni} 
\affiliation{Dipartimento di Fisica, Università di Pisa, Largo Bruno Pontecorvo 3, 56127, Pisa, Italy } 
\affiliation{INFN, Largo B. Pontecorvo 3, 56127, Pisa, Italy} 
\affiliation{INAF - Osservatorio di Astrofisica e Scienza dello Spazio di Bologna, Via Piero Gobetti 93/3, 40129, Bologna, Italy}
\email{michele.cignoni@unipi.it}

\author[0000-0002-1723-6330] {B.\, G.\, Elmegreen}
\affiliation{Katonah, NY 10536 USA}
\email{belmegreen@gmail.com}

\author[0000-0002-1392-3520] {D.\, M.\, Elmegreen}
\affiliation{Dept. of Physics \& Astronomy, Vassar College, Poughkeepsie, NY 12604 USA}
\email{elmegreen@vassar.edu}

\author[0000-0001-8608-0408] {J.\, S.\, Gallagher III}
\affiliation{Department of Astronomy, University of Wisconsin-Madison, Madison, WI 53706 USA}
\email{jsg@astro.wisc.edu}

\author[0000-0002-5581-2896]{M.\, Gennaro}
\affiliation{Space Telescope Science Institute, 3700 San Martin Dr, Baltimore, MD 21218, USA}
\affiliation{The William H. Miller {\sc III} Department of Physics \& Astronomy, Bloomberg Center for Physics and Astronomy, Johns Hopkins University, 3400 N. Charles Street, Baltimore, MD 21218, USA}
\email{gennaro@stsci.edu}

\author[0000-0002-1891-3794]{E.\ K.\, Grebel}
\affiliation{Astronomisches Rechen-Institut, Zentrum f\"ur Astronomie der Universit\"at Heidelberg, M\"onchhofstr. 12-14, D-69120 Heidelberg, Germany}
\email{grebel@ari.uni-heidelberg.de}

\author[0000-0002-0560-3172]{R.\ S.\ Klessen}
\affiliation{Universit\"{a}t Heidelberg, Zentrum f\"{u}r Astronomie, Institut f\"{u}r Theoretische Astrophysik, Albert-Ueberle-Str.\ 2, 69120 Heidelberg, Germany}
\affiliation{Universit\"{a}t Heidelberg, Interdisziplin\"{a}res Zentrum f\"{u}r Wissenschaftliches Rechnen, Im Neuenheimer Feld 225, 69120 Heidelberg, Germany}
\affiliation{Harvard-Smithsonian Center for Astrophysics, 60 Garden Street, Cambridge, MA 02138, U.S.A.}
\affiliation{Radcliffe Institute for Advanced Studies at Harvard University, 10 Garden Street, Cambridge, MA 02138, U.S.A.}
\email{klessen@uni-heidelberg.de}

\author[0000-0001-5171-5629]{A.\, Pasquali}
\affiliation{Astronomisches Rechen-Institut, Zentrum f\"ur Astronomie der Universit\"at Heidelberg, M\"onchhofstr. 12-14, D-69120 Heidelberg, Germany}
\email{pasquali@ari.uni-heidelberg.de}

\author[0000-0002-0806-168X]{L.\, J.\, Smith}
\affiliation{Space Telescope Science Institute, 3700 San Martin Dr, Baltimore, MD 21218, USA}
\email{lsmith@stsci.edu}

\author[0000-0001-8289-3428]{A.\, Wofford}
\affiliation{Universidad Nacional Aut{\'o}noma de M{\'e}xico, Instituto de Astronom{\'i}a, AP 106, Ensenada 22800, BC, M{\'e}xico}
\email{awofford@astro.unam.mx}

\begin{abstract}

The Galaxy UV Legacy Project (GULP) is a Cycle 28 Treasury program with the Hubble Space Telescope (HST) designed to characterize resolved massive stars, OB associations, and young star clusters (YSCs) in 26 nearby star-forming galaxies. Utilizing the ACS/SBC F150LP and WFC3/UVIS F218W filters, combined with extensive archival observations, GULP provides an unprecedented panchromatic 8-band view from the Far-UV to the I-band. The target galaxies were carefully selected to span a broad range of metallicities, masses, morphological types, and star formation rates, thereby enabling detailed studies of star formation processes across different galactic environments.
This paper introduces the GULP survey, detailing its observational strategy, data processing, and initial scientific results for the irregular barred starburst dwarf galaxy NGC 4449, used as a test case. We derived the physical parameters and ages for thousands of stars using the Binary Populations And Spectral Synthesis (BPASS) models, and found that the younger stars and clusters are predominantly concentrated along the galaxy's central bar, and that over the past $<50\, {\mathrm{Myr}}$ star formation progressively migrated from northeast to southwest.
We used the F150LP, F218W, and F275W filters to investigate how the ``UV-bump'' at $\lambda 2175\, \mathrm{\text{\AA}}$ correlates with the intensity of the UV radiation. The ``UV-bump'' is detected in many areas of the galaxy, but is absent in the regions of most intense and recent star formation. This strongly supports the scenario where UV radiation from young, massive stars effectively destroys the small dust grains responsible for the ``UV-bump''.

\end{abstract}

\section{Introduction} 
\label{sec:intro}

\begin{figure} \centering
\includegraphics[width=\columnwidth]{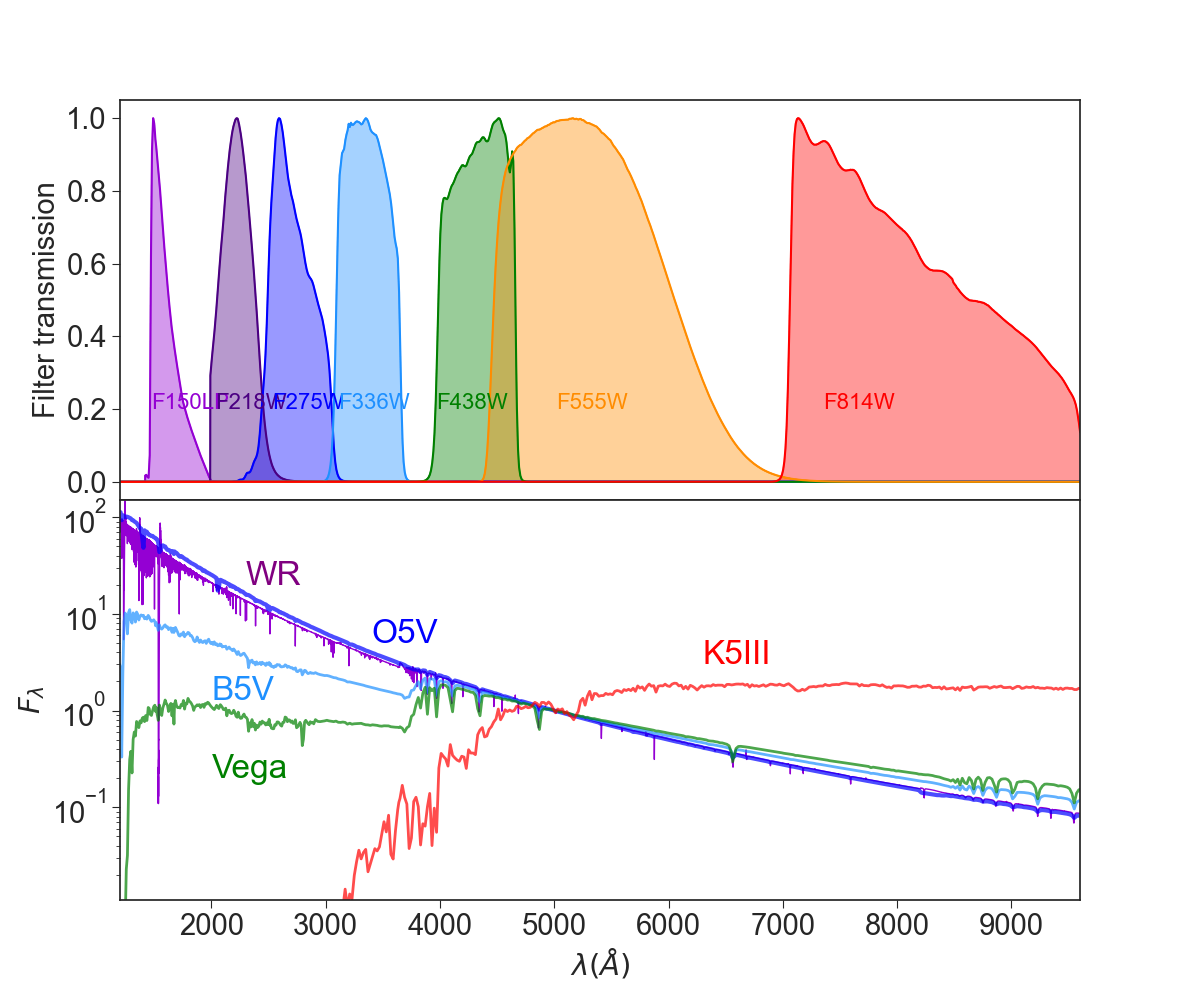} 
\caption{Upper panel: Relative transmission of the broadband filters used in the GULP survey as a function of wavelength. Lower panel: Atmosphere models for stars with different temperatures and surface gravities, including a WR star \citep[purple; ][]{Hamann2004}, an O5V star (dark blue), a B5V star (cyan), Vega (green), and a K5\textsc{iii} red supergiant (red). Apart from the WR star, the remaining models are from \citet{Kurucz1993}. Fluxes are normalized at $\lambda = 5000\, {\textrm{\AA}}$.}
\label{f:GULPfilters}
\end{figure}

Massive O-type stars ($M\geq 16\, \mathrm{M_\odot}$) are short-lived ($\lesssim 10\, \mathrm{Myr}$). While in the Kroupa initial mass function \citep[IMF, ][]{Kroupa1993} they represent less than 0.1\% of all stars between $0.1-120\, \rm{M}_\odot$, they account for over 10\% of the stellar population's total mass, and, with their UV radiation, stellar winds, and explosive ending, they inject energy, momentum, and chemical elements into the interstellar (ISM) and intergalactic medium (IGM). This process has been regulating the galaxy stellar mass growth since the formation of the first stars \citep[e.g.][]{Kennicutt2005, Shepherd1996a, Shepherd1996b, MacLow2004, Urquhart2014}. 

Our knowledge of the properties of O-type stars is largely confined to the Milky Way (MW) and its nearby satellites. Large distances, combined with the need for space-based UV capabilities, have limited our ability to explore how local and global conditions influence the formation and evolution of these objects, and to quantify their effects on surrounding environments.

The Director's Discretionary program Ultraviolet Legacy Library of Young Stars as Essential Standards \citep[ULLYSES]{Roman-Duval2020, Roman-Duval2025} collected UV spectra of about 240 OB stars in the Magellanic Clouds (MCs) and six fainter stars in the metal-poor Local Group (LG) galaxies NGC~3109 and Sextans~A with the Hubble Space Telescope (HST) Cosmic Origin Spectrograph (COS) and Space Telescope Imaging Spectrograph (STIS). In addition, over the past two decades, several HST surveys have been exploring the processes that trigger, regulate, and quench the formation of massive stars and clusters in the LG. ANGST \citep[PI Dalcanton, GO-10915,][]{Dalcanton2009}, PHAT \citep[PI Dalcanton, GO-12055,][]{Dalcanton2012}, and PHATTER \citep[PI Dalcanton, GO-14610,][]{Williams2021} studied the properties  of massive stars and young clusters in the U-band ($300 \lesssim \lambda \lesssim 400\, \mathrm{nm}$) and near-ultraviolet (NUV) ($230 \lesssim \lambda \lesssim300\, \mathrm{nm}$). LEGUS \citep[PI Calzetti, GO-13364,][]{Calzetti2015, Adamo2017, Sabbi2018}, Hi-PEEC \citep[PI Adamo, GO-14066][]{Adamo2020}, and PHANGS-HST \citep[PI Lee, GO-15654][]{Lee2022} further broadened the sampled properties by including star-forming galaxies in the local universe. 

The large HST program GO-11079 \citep[PI Bianchi, ][]{Bianchi2012, Bianchi2014} provided the first high-spatial-resolution far-ultraviolet (FUV) observations of six LG galaxies using the Wide Field Planetary Camera 2 (WFPC2). The Galaxy UV Legacy Project (GULP, PI Sabbi, GO-16316) now significantly expands upon this work by adding high-spatial-resolution, wide-field coverage HST observations in the FUV using the Solar Blind Channel (SBC) filter F150LP ($\mathrm{FWHM_{PSF}} = 0\farcs{06},\, 140 \lesssim \lambda \lesssim 190\, \mathrm{nm}$) of the Advanced Camera for Surveys (ACS) and in the NUV using the UVIS Channel filter F218W ($\mathrm{FWHM_{PSF}} = 0\farcs{08},\, 199 \lesssim \lambda \lesssim 244\ , \mathrm{nm}$) of the Wide Field Camera 3 (WFC3) for 26 nearby star-forming galaxies, already observed with HST at NUV, U and optical wavelengths. In Figure~\ref{f:GULPfilters} we show the transmission curves of the filters used in GULP (including archival observations) and how they compare to the stellar spectral types of stars in different evolutionary phases.

The paper is organized as follow: Section \ref{sec:sample_sel} describes the selection criteria used to choose the targets. Section \ref{sec:obs} summarizes the observing strategy and the analysis of the data. Section~\ref{sec:CMDs} discusses the properties of NGC~4449 resolved stellar populations, Section~\ref{sec:YSCs} presents the characteristics of NGC~4449 young star clusters (YSCs), and Section~\ref{sec:SFH} discusses the spatial and temporal progression of star formation in NGC~4449. The galaxy dust properties are shown in Section~\ref{sec:UV-bump}. Our conclusions are presented in Section~\ref{sec:conclusions}.

\section{GULP's Sample Selection}
\label{sec:sample_sel}

\begin{figure*}
\plotone{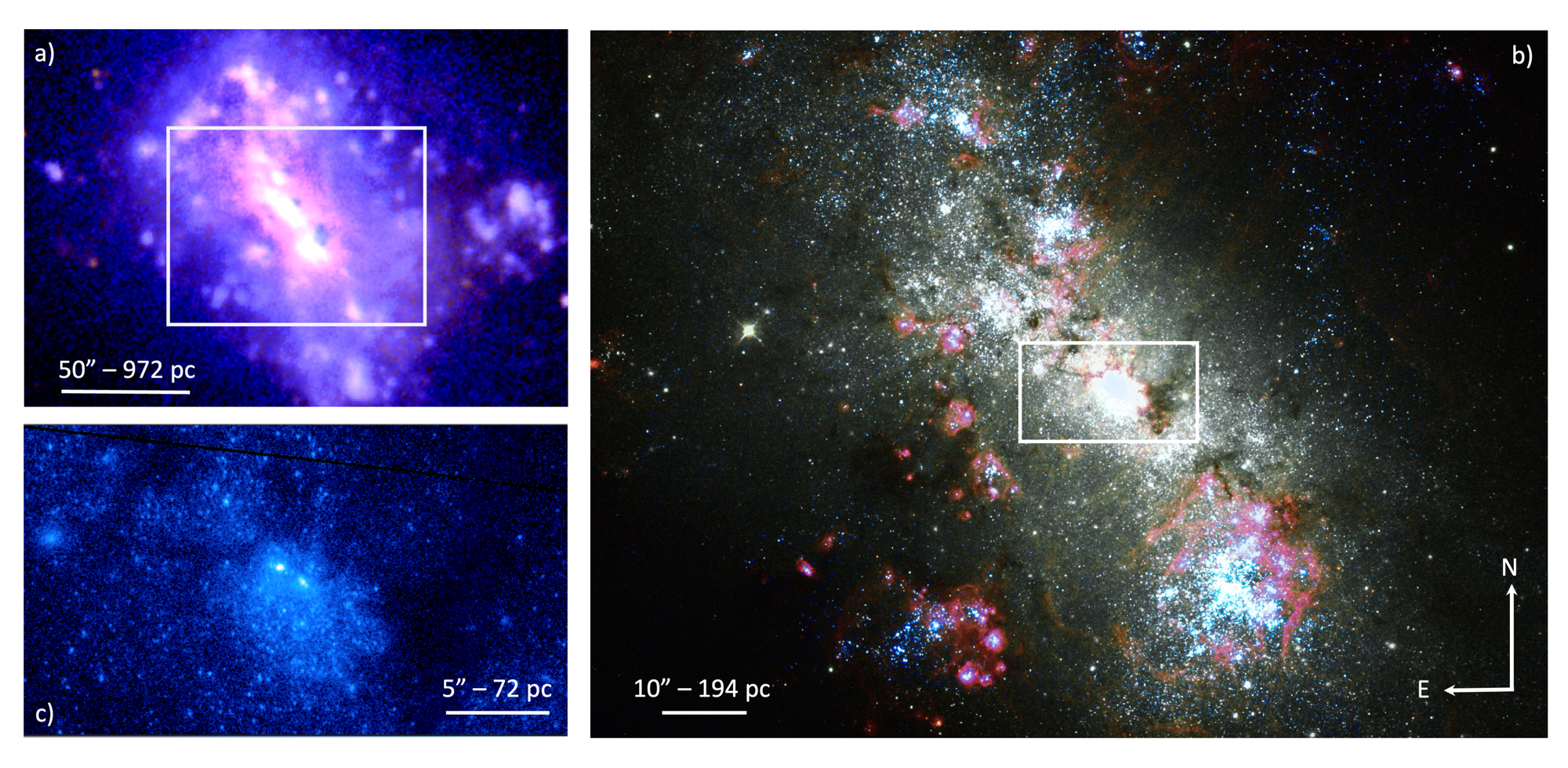}
\caption{The Magellanic-type galaxy NGC~4449. Panel a) shows the GALEX color-composite image with the FUV filter in blue and the NUV filter in red. The white rectangle highlights the region covered by GULP with the SBC filter F150LP. Panel b) shows the color composite image of the portion of galaxy studied with the SBC. The filter F150LP is depicted in blue, the ACS/F658N filter in red, and ACS/F438W in green. Panel c) presents a $\sim 447\times 277\,\mathrm{pc}$ zoom-in of F150LP image of the central double YSC. The corresponding region  in panel b) is marked with a white rectangle. In all three panels, North is up, and East is toward the left. 
}
\label{f:GULPvsGALEX}
\end{figure*}

\begin{figure*}
\includegraphics[width=1\textwidth, trim=0 20 0 70, clip]{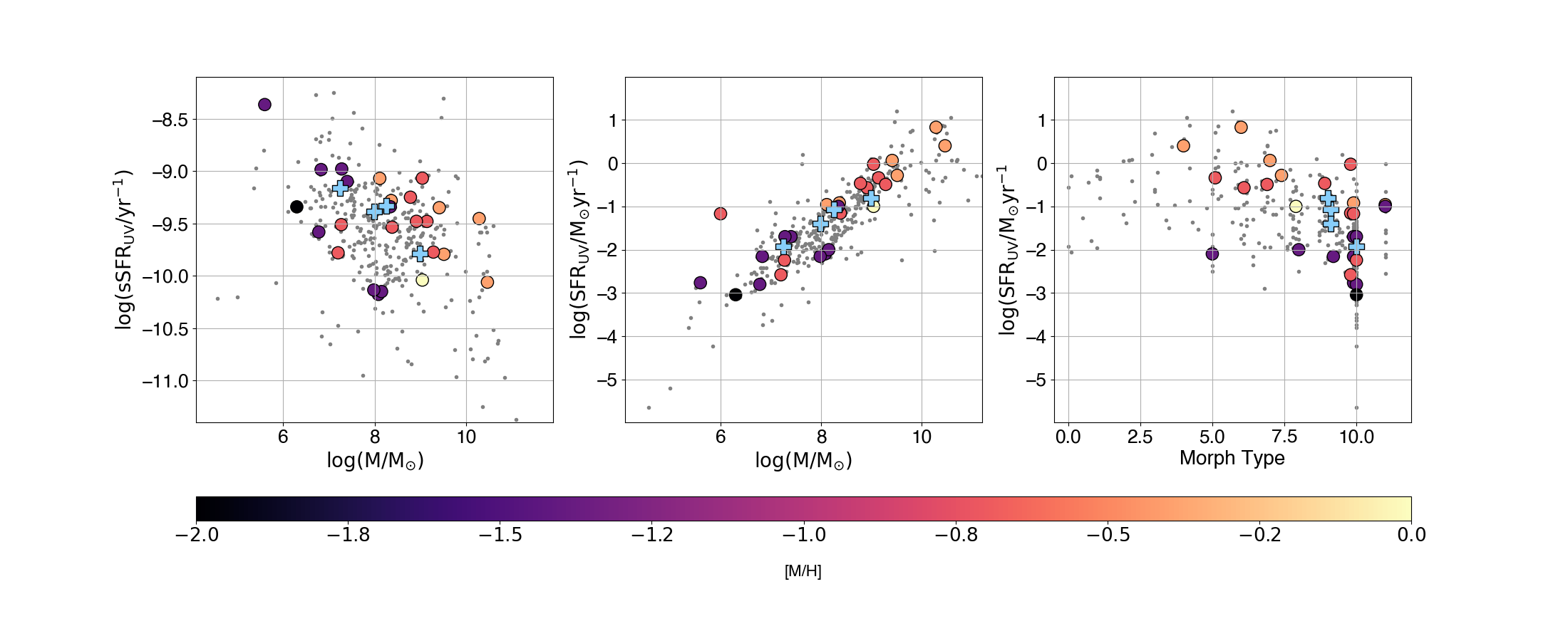}
\caption{Left Panel: the sSFR and mass, both in logarithmic scale, for 276 star-forming galaxies found within $8\, \mathrm{Mpc}$ \citep[gray points,][]{Kennicutt2008}. GULP galaxies are color-coded from dark purple to light yellow, for increasing metallicity. The light blue crosses represent the four galaxies (the 2 MCs, NGC~3109, and Sextans~A) observed by  ULLYSES. Central Panel: The same as the left panel, but for the logarithm of the SFR. Right Panel: SFR, on logarithmic scale, as a function of the morphological types. SFRs are derived from GALEX UV luminosity \citep{Lee2009}, scaled to the area covered by the F275W filter.}
\label{f:sample}
\end{figure*}

\begin{figure*}
\includegraphics[width=1\textwidth]{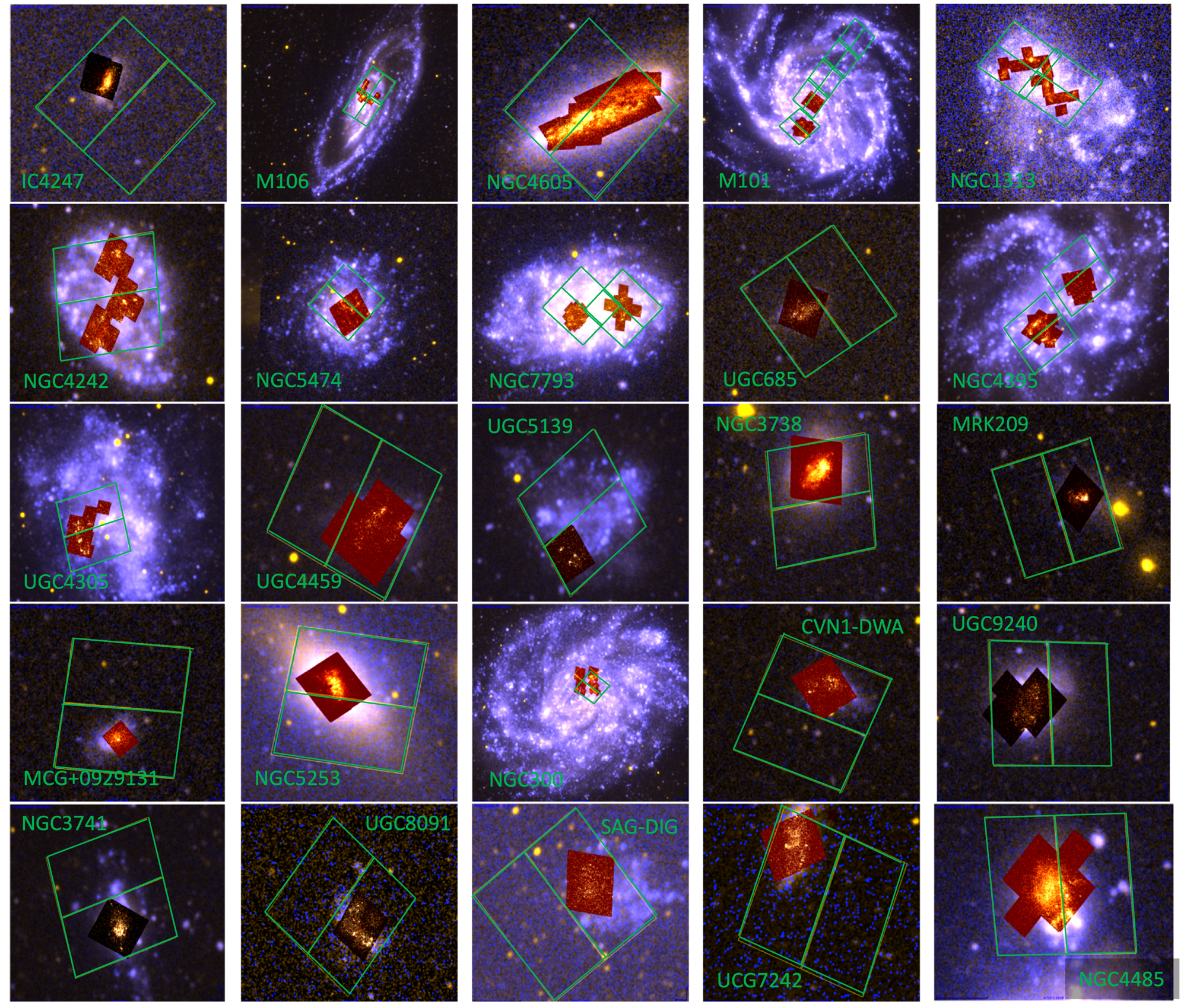}
\caption{{Footprints of the GULP observations, except NGC~4449, already shown in Fig.~\ref{f:GULPvsGALEX}. The ACS/SBC mosaics, in blackbody Asinh color-map}, are superimposed over the GALEX color-composite images (the FUV chanel is in blue and the NUV chanel in red). The $2\farcs7 \times 2\farcs7$ footprint of the archival UVIS observations in the F275W filter, used to target the brightest UV regions that do not saturate the SBC detector, is shown in green. }
\label{f:targets}
\end{figure*}

GULP is a large FUV and NUV treasury program designed to characterize the properties of resolved massive stars, OB associations, and YSCs in nearby ($<8\,\mathrm{ Mpc}$) star-forming galaxies. This  efficient survey leverages HST ACS and WFC3 observations already available in the Barbara A. Mikulski Archive for Space Telescope (\href{https://mast.stsci.edu/portal/Mashup/Clients/Mast/Portal.html}{MAST}) at wavelengths longer than $\lambda> 2930\, \mathrm{\text{\AA}}$, often including H$\alpha$. 

Massive stars are preferentially found in compact, dense groups. With a $\sim 60-80$ times better angular resolution than GALEX ($\mathrm{FWHM} = 0\farcs06$ vs. $4\farcs2$ in the FUV, and $0\farcs08$ vs $5\farcs0$ in the NUV, see also  Fig.~\ref{f:GULPvsGALEX}), GULP provides, for the first time, spectral energy distributions (SEDs) between $\sim 1500$ {\AA} and $\sim 8200$ {\AA} for hundreds of thousands of OB stars and YSC, and enables several otherwise unattainable scientific results, such as:
\begin{itemize}
\item Improve the age (and therefore mass) estimates of star clusters and OB associations; 
\item Investigate the high-stellar mass regime of the IMF; 
\item Constrain the prevalence of binary stars in the high-mass regime;
\item Characterize the clustering properties of star-forming regions \citep[e.g.,][]{Meena2025}; 
\item Investigate the origin of isolated massive stars \citep[e.g., ][]{Facchini2025};
\item Determine the size of dust grains in different local environments;
\item Quantify the ionizing photon production rate;
\item Provide empirical templates to calibrate the models used to interpret the light from galaxies observed at higher redshifts. 
\end{itemize}

GULP targets 26 nearby star-forming galaxies, carefully selected from the $11\, {\mathrm{Mpc}}$ Ultraviolet and H$\alpha$ Survey \citep[11HUGS,][]{Kennicutt2008} catalog, to span as evenly as possible a broad range of galactic properties and morphologies, and enable detailed studies of resolved SF and YSC evolution. Sampling a wide range of metallicity values is crucial because of the impact on the extinction law, observed SEDs, stellar wind mass loss rates, and feedback from SF \citep{Vink2001, Kudritzki2002,Mokiem2007, Ramachandran2018, Ramachandran2019}. Similarly, sampling a variety of SFRs is important because of the influence that SFR has in the clustering fraction \citep{Kruijssen2012, Adamo2015, Johnson2017, Grasha2019}. To build our sample, we thus considered the five-dimensional parameter space (see Fig.~\ref{f:sample}), including: SFR  ($-3 \le \log(\mathrm{SFR/(M_\odot yr^{-1}})) < 1$), specific star formation rate ($\mathrm{sSFR} = \mathrm{SFR}/\mathrm{M_\odot}$; $-11.5<\
\log(\mathrm{sSFR/yr^{-1}})<-8.5$), metallicity ($-2.1<[{\mathrm{M/H}}]<0$ dex), major morphological types (Sa, Sb, Sc, Sd, Sm, Irr), and mass ($5< \log(\mathrm{M/M_\odot})<11$). The SFRs used for the selections were obtained from UV GALEX observations \citep{Lee2009}, scaled to the surveyed area. 

The list of galaxies, along with the used filters and a summary of their properties, such as distance, \citep[derived from the tip of the red giant branch by][]{Sabbi2018}, morphology, UV SFR, mass, and metallicity \citep[from][]{Calzetti2015}, is provided in Table~\ref{t:targets}. Figure~\ref{f:targets} shows the area mapped by GULP for each target, where the SBC mosaics are superimposed on the GALEX images. The footprint of the archival UVIS image in the filter F275W, used to identify the regions with higher UV emission, is shown in green.

\section{Observations}
\label{sec:obs}

The GULP survey was awarded 84 orbits of HST time to observe 26 nearby (between 2 and $8\, {\mathrm{Mpc}}$) star-forming galaxies using the FUV/SBC filter F150LP and NUV/UVIS filter F218W. The F218W filter was selected because it is sensitive to the strength of the ``UV-bump'' -- an extinction maximum at $\lambda = 217.5 \pm 1\, \mathrm{nm}$ \citep{Stecher1965a}. Observations in the F218W filter can therefore be used to break the degeneracy between effective temperature and reddening. Conversely, the F150LP filter was chosen to constrain the temperature and age of massive stars and young star clusters. 

Each of the GULP targets had already been observed in at least two broad-band imaging filters, either with UVIS or the Wide Field Channel (WFC) of ACS. Specifically, five galaxies had been previously observed in F435W, F555W, and F814W by ANGST \citep{Dalcanton2009}, while the remaining 21 galaxies had been observed in five broad-band filters (F275W, F336W, F438W, F555W, and F814W) by LEGUS \citep{Calzetti2015}.

To better constrain the age of the YSCs and quantify the amount of feedback from massive stars, we also requested H$\alpha$ observations. Observations  from various programs were already available for 16 galaxies (in the filters F656N, F657N, or F658N). For the remaining ten galaxies, we acquired 3-point dithered observations with the UVIS F657N filter. Where not previously available, we included the UVIS filters F275W, F336W, F438W, F555W, and/or F814W to ensure each galaxy had 8-filter coverage (typically F150LP, F218W, F275W, F336W, F438W, F555W, F657N, and F814W) from the far-UV to the I band. Note that for NGC 300, we used the already available WFC3 filter F225W instead of F275W. The entire source dataset can be found in MAST using the digital object identifier (DOI): \dataset[doi:10.17909/z182-0m23]{https://doi.org/10.17909/z182-0m23}, or under the dataset name ``GULP: Galaxy UV Legacy Project''.

Information on the GULP observations is reported in Table \ref{t:targets}. This includes:
\begin{itemize}
\item The number of pointings used for the F150LP and F218W filters.
\item Which WFC3 filters (other than F218W) were acquired by GULP.
\item Which archival ACS/WFC filters were used as substitutes for the UVIS filters F438W, F555W, or F814W.
\end{itemize}

The exposure times in the F150LP filter were tailored to observe a $16\, {\mathrm{M_\odot}}$ star with ${\mathrm{S/N}} > 6$ in all galaxies. For galaxies closer than $5\, {\mathrm{Mpc}}$, this S/N can be achieved in $\sim 230\, {\mathrm{s}}$. For more distant galaxies similar performances can be reached in $690\, {\mathrm{s}}$.   

The SBC combines high spatial resolution ($\mathrm{FWHM_{PSF}} = 0\farcs06$, corresponding to 0.6 and $2.3\, {\mathrm{pc}}$ at a distance of 2 and $7.5\, {\mathrm{Mpc}}$, respectively) with FUV sensitivity (wavelength range between 142 and $199\, \mathrm{nm}$) at the cost of a very small field of view ($\sim 34\farcs6 \times 30\farcs0$). To overcome this limitation, we created mosaics consisting of 8 or 9 adjacent $230\, {\mathrm{s}}$ exposures. The resulting mosaics were carefully designed to avoid guide star re-acquisitions and memory dumps, ensuring that the observations of the closest galaxies fit within a single orbit. For the furthest galaxies we used 3 orbits. 

Within the distance range of 2 to $7.5\, \mathrm{Mpc}$, a $3\times 3$ ACS/SBC mosaic subtends 1.3 to $5.1\, \mathrm{kpc}$, covering, in several cases, a major fraction of the galaxy's star-forming disk (see Fig.~\ref{f:targets}). Thus, by restricting the distance range covered by our survey to $< 8.0\, \mathrm{Mpc}$, we achieve an optimal compromise between spatial resolution and coverage. {The three major spiral galaxies M101, M106, and NGC~300 are a noticeable exception. Given their size, for M101 and M106 we decided to probe how the younger stellar populations vary radially. For NGC~300, we leveraged the FUV observations already available in the archive, adding only the ``UV-bump'' sensitive filter F218W. As a result, only a fraction of NGC300 is covered; nevertheless the data allow us to explore both the arm and inter-arm stellar populations.}

We used the archival F275W UVIS observations to further customize the shape of the SBC mosaics to the brightest UV regions (see Fig.~\ref{f:targets}). Additionally, the F275W observations allowed us to confirm that none of our pointings exceeded the SBC's local and global count rate safety limits.

\subsection{Imaging and Mosaics}
\label{sec:mosaics}

The entire dataset was processed with the standard Space Telescope Science Institute calibration pipelines, CALACS version 10.3.5 and CALWFC3 version 3.6.2. The pipelines subtract bias level, superbias and superdark, and apply the flat-field correction. For the ACS WFC and WFC3 UVIS observations, the pipelines apply also the time-dependent pixel-based CTE correction to the flat-fielded images. No CTE correction is needed for the ACS SBC observations. 

We aligned the resulting F438W (or F435W) ``{\tt \_flc}'' exposures to Gaia Data Release 3 \citep[DR3,][]{Gaia2021} using the Python routine {\tt TweakReg} in the {\tt DrizzlePac} package version 3.5.1 to build the astrometric reference frame of each target. The shift, scale, and rotation of the individual exposures were solved using catalog matching to an accuracy of better than 0.1 pixels. The {\tt DrizzlePac} routine {\tt AstroDrizzle} was then used to combine the aligned images into a ``{\tt \_drc}'' mosaic with a pixel scale of $39.62\, \mathrm{mas\, pixel^{-1}}$ and to update the coordinate system in the headers of the individual exposures.

We used the F438W (or F435W) mosaics to align the exposures of the remaining seven filters. All the mosaics are in $\mathrm{electrons\, s^{-1}}$, which enables a user to convert instrumental measurements to magnitudes using the time-dependent photometric zero points stored in the header of the images. We note that our mosaics differ from those released by the LEGUS team \citep{Calzetti2015, Sabbi2018} both in terms of absolute astrometry and for photometric zero points. These differences are due to the fact the LEGUS exposures are aligned using the GSC2 astrometry system, while our data are aligned to Gaia DR3 \citep{Gaia2023}, and the WFC3 Team released updated flat fields and new time-dependent zero points about one year after the release of the LEGUS high-level science products (HLSP). We will release the updated single exposures and the mosaics to MAST in the GULP HLSP \dataset[doi:10.17909/mgsc-0985]{https://doi.org/10.17909/mgsc-0985}.

\subsection{Stellar Photometry}
\label{sec:stellar_phot}

\begin{figure*}
\includegraphics[width=1\textwidth]{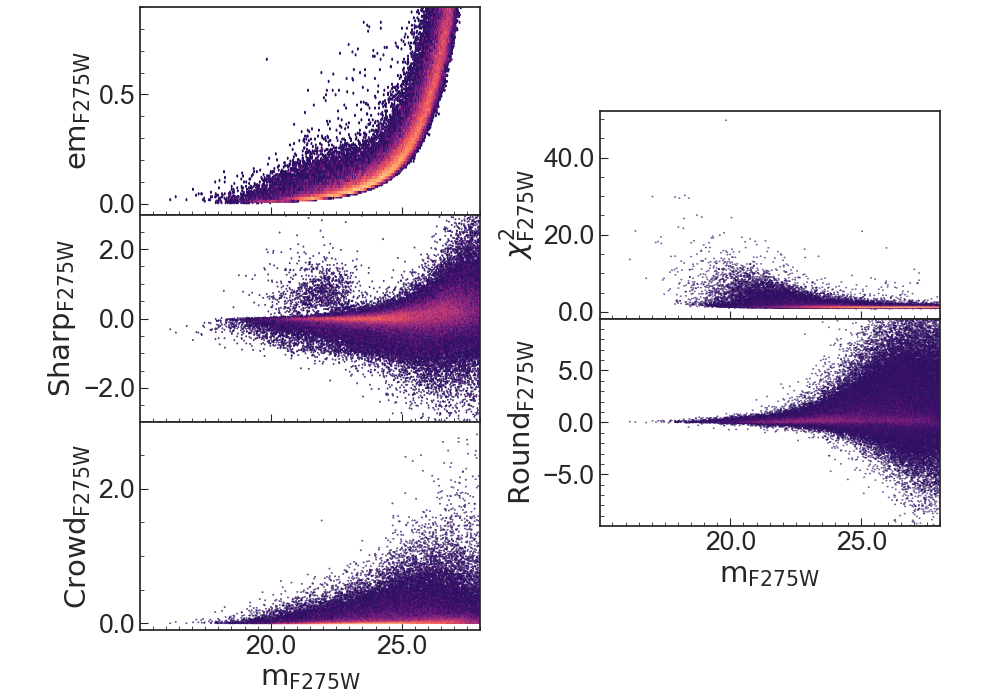}
\caption{NGC~4449 DOLPHOT photometry errors (upper-left panel), $\chi^2$ (upper-right panel), sharpness (middle-left panel), PSF-roundness (middle-right panel), and crowding (lower-left panel) plotted as a function of $\rm{m}_{F275W}$.}
\label{f:dolphot_phot}
\end{figure*}

Positions and fluxes for point-like sources in the F218W, F275W, F336W, F438W(F435W), F555W, and F814W filters were measured via PSF-fitting using the WFC3 and ACS modules of the photometry package {\tt DOLPHOT} version 2.0, downloaded on January 2021 from the website \url{http://americano.dolphinsim.com/dolphot/} \citep{Dolphin2000, Dolphin2016}. All filters were analyzed together, using the F438W (or F435W) drizzled mosaic as reference.

{\tt DOLPHOT} carries out the photometry independently in each filter. It iteratively identifies peaks above the detection threshold and uses PSF libraries to fit simultaneously the PSF and the sky to every peak within a group of images. In our analysis we used the PSF libraries with Jay Anderson's core (see DOLPHOT manual). In each filter, we set the initial detection threshold on the individual images to {\tt SigFind} $> 5.5\, \sigma$. We then retained all the sources whose weighted average magnitude was {\tt SigFinal} $> 3.5\, \sigma$ above the average measurements in the individual exposures. 

In each filter, {\tt DOLPHOT} identifies isolated stars to determine small aperture correction adjustments to account for minor differences between the PSF libraries and the observed PSF, caused, for example, by the telescope's breathing. The final measured count rates are then converted into the {\sc Vegamag} system by adding hard-coded zero points. Due to flat-field revisions performed over the years by the WFC3 team \citep[to better reflect the differences in quantum efficiency between UVIS' two chips,][] {Ryan2016}, and a slow but progressive decrease in the sensitivity of both ACS and WFC3 detectors \citep{Bohlin2016, Calamida2022}, {\tt DOLPHOT}'s zero points are now out of sync with the most up-to-date calibrations. To account for this effect, we subtracted {\sc dolphot}'s hard-coded zero points before applying the appropriate ones, obtained from the {\tt photlam} keyword in the image headers. 

For each star, DOLPHOT provides several parameters, in addition to magnitudes, to assess the quality of the photometry. These include: photometry errors, estimates of the quality of the PSF-fitting (expressed through the $\chi^2$ value of the residuals), sharpness, roundness, contamination from nearby sources (crowding value), object type classification, and a photometry quality number. The distribution of the first five parameters for the filter F275W is shown in Fig.~\ref{f:dolphot_phot} as an example. Additionally, in our catalogs we included the number of exposures that contributed to each measurement, the corresponding total exposure time (as derived by the number of exposures), and the resulting SNR.

{\tt DOLPHOT} does not support the SBC observations, and, because of the small number of individual stars, it failed to analyze the narrow-band H$\alpha$ exposures. Given the relatively low stellar density in these two filters, we performed a simple aperture photometry using the Astropy package {\tt Photutils-1.8.0} \citep{Bradley2023}, with aperture radius $r_a = 3\, \mathrm{pixels}$, and an {annulus with inner radius of $r_i = 10$ and width of $2,\, \mathrm{pixels}$ for sky subtraction}. To minimize the number of spurious detections, we used as input list the stars detected by {\tt DOLPHOT} in the F275W filter. 

The comparison between the SNR and the photometry errors indicates that {\tt Photutils} significantly overestimates the latter, providing an error of $0.1\, \mathrm{mag}$ for sources with $\mathrm{SNR}=100$, and $0.5\, \mathrm{mag}$ for $\mathrm{SNR}=20$. Therefore, for these two filters, we decided to rely on the SNR and  defined the magnitude error as:
$
{\mathrm{MAGER}} = 0.5*(|-2.5*\log(1+1/{\mathrm{SNR}})|+|-2.5*\log(1-1/{\mathrm{SNR}})|)
$. {The obtained values should be considered lower estimates. While they provide a good measure of the shot noise associated with a source, they do not account for other uncertainty sources, such as readout noise or errors in the background estimation.}

The resulting eight-band photometry catalogs (in the case of NGC~4449: F150LP, F218W, F275W, F336W, F435W, F555W, F658N, F814W) will be available for download through MAST at the GULP HLSP: \dataset[doi:10.17909/mgsc-0985]{https://doi.org/10.17909/mgsc-0985}. In addition to photometry information, the catalogs include a unique stellar identifier ``GULP+RaDec'', the astrometric coordinates Ra and Dec (J2016), and the pixel coordinates on the reference mosaic. 

\subsection{Star Clusters Photometry}
\label{sec:cluster_phot}

With the exception of the dwarf galaxies, characterized by modest SFRs, most of the galaxies in GULP host rich populations of young, intermediate and old star clusters.

We used the catalogs of cluster candidates released by the LEGUS team\footnote{https://legus.stsci.edu/} as input for our analysis.  The sources were automatically identified using a custom pipeline (legus\_clusters\_extraction.py 4.0) and then vetted by eye by a few LEGUS team members. On the basis of contour and surface plots, each source was classified either as Class~1 (= compact and centrally concentrated, with FWHM more extended than for typical stars), Class~2 (= slightly elongated density profiles and less symmetric light distribution), and Class~3 (= asymmetric profiles and multiple peaks on top of diffuse underlying wings, which suggest the presence of a possible concentration of low mass stars) objects. Any interlopers (stars, background galaxies, or image artifacts) that survived the automated cluster filtering were flagged as Class~4 \citep[see ][for a more detailed discussion of the cluster selection and classification]{Adamo2017}. 

To measure the clusters magnitudes in the filters F150LP and F218W, and to update their magnitudes in the archival filters, we adopeted a modified version of the recently developed {\tt FEAST-pipeline}. The pipeline will be described in detail in Adamo et al. (in prep.); here we focus on the main steps taken to produce our photometric catalogs. 

Since the photometric data are all aligned, we allowed for improved centering of each source in the reference filter (F435W in the case of NGC~4449). In each filter, we measured the cluster magnitude within apertures of radii 1, 3, and 5 pixels. The local background contribution was determined on a 2 pixel-wide sky annulus located at a 7 pixel radius. {We defined the difference in magnitude between the 1 and 3 pixel apertures as the filter-based concentration index (CI) for each cluster.

Studies of star cluster populations in nearby galaxies (within 10 Mpc) indicate that the surface brightness profiles of YSCs are well-described by a Moffat function with power-law index $\beta=1.5$ and an average effective radius $R_{eff}\sim 2 - 3\, {\rm pc}$ \citep{Elson1987, Larsen2004, Ryon2015}. \citet{Ryon2017} found a tight correlation between the cluster’s  $R_{eff}$ and its CI. Following the methodology of the LEGUS and FEAST teams \citep[e.g., ][]{Ryon2017, Knutas2025}, we created a grid of Moffat models with fixed index $\beta=1.5$ and radii ranging from 0.3 and 5 pc, converted in angular scales based on the adopted distance of the galaxy. We then convolved these models with the stellar PSF of the reference image, and derived CI for each model in the same manner as the observations. We used the CI--$R_{eff}$ relation to associate the observed CI of each source to its expected $R_{eff}$, thus identifying the closest Moffat model that describes the source light-profile to estimate the aperture correction at 20 pixels. We repeated the same procedure for all the other filters.}

\section{NGC~4449's Stellar Populations}
\label{sec:CMDs}

\begin{figure*}
\includegraphics[width=1\textwidth]{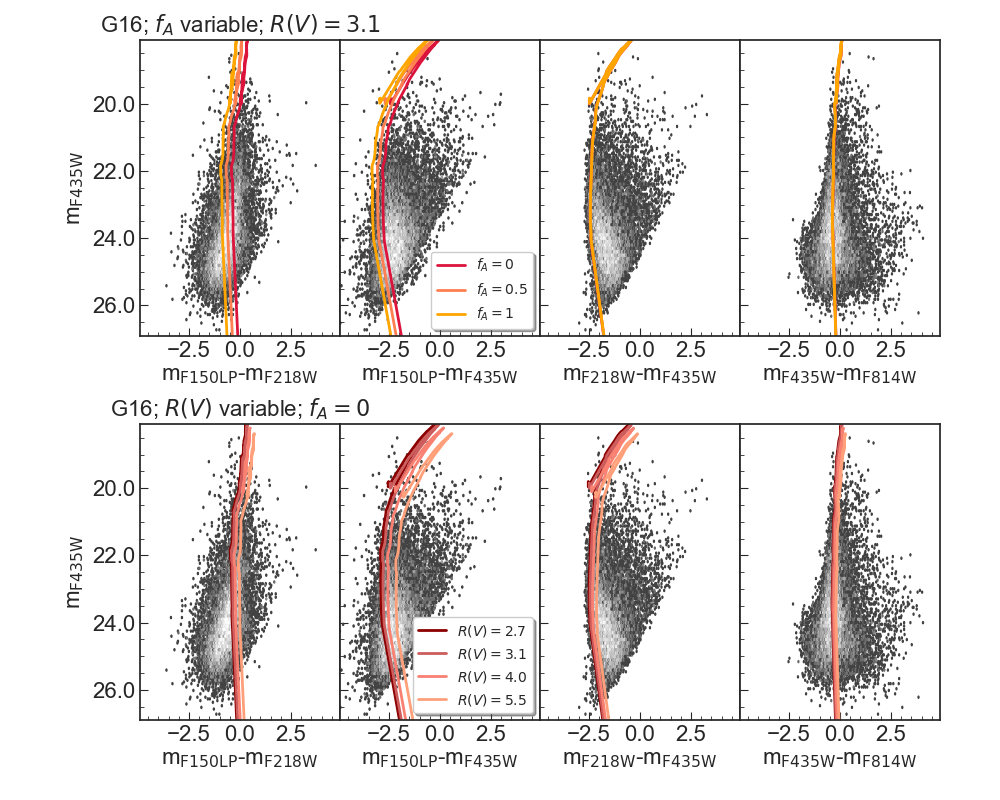}
\caption{NGC~4449 Hess diagrams in the filters  $\rm{m}_{F435W}$ vs. $\rm{m}_{F150LP}-\rm{m}_{F218W}$, $\rm{m}_{F150LP}-\rm{m}_{F435W}$, $\rm{m}_{F218W}-\rm{m}_{F435W}$, $\rm{m}_{F435W}-\rm{m}_{F814W}$. Only stars detected in all seven broadband filters with photometric error $<0.5\, {\mathrm{mag}}$ are plotted. A 3 Myr Padova isochrone (metallicity Z=0.004, distance modulus DM = 28.02, and $E(B-V)=0.07$) is shown using the \citet{Gordon2016, Gordon2024} two-parameter formalism of the extinction law. In the upper panels, the total-to-selective extinction ratio $R(V)$ is set to 3.1 and mixing parameter $f_A$ varies from 0 (in red), 0.5 (in orange) to 1.0 (in yellow). In the lower panels, $f_A$ is set to 0, while $R(V)$ ranges from 2.7 (dark red), 3.1 (red), 4.0 (salmon), to 5.5 (light salmon).}
\label{f:extinction_laws}
\end{figure*}

\begin{figure*}
\includegraphics[width=1\textwidth]{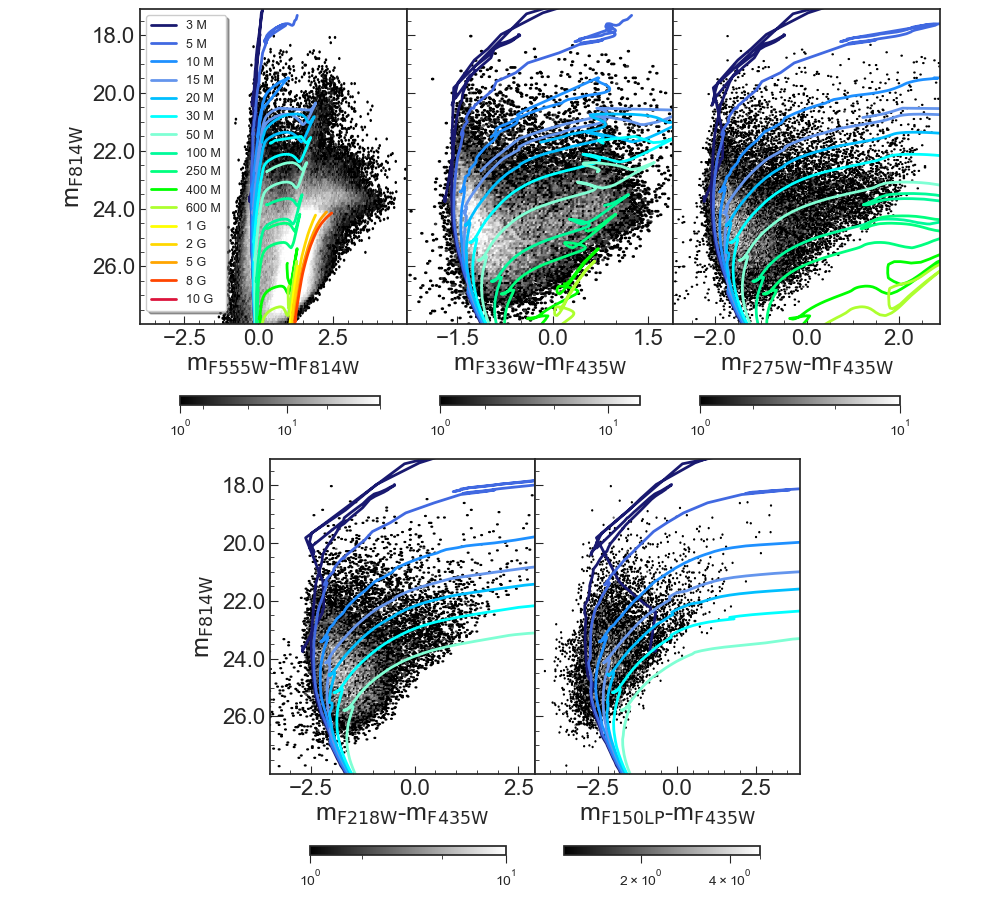}
\caption{From top to bottom, and from left to right, NGC~4449 Hess diagrams in the filters $\rm{m}_{F814W}$ vs. $\rm{m}_{F555W}-\rm{m}_{F814W}$, $\rm{m}_{F336W}-\rm{m}_{F435W}$, $\rm{m}_{F275W}-\rm{m}_{F435W}$, $\rm{m}_{F218W}-\rm{m}_{F435W}$, and $\rm{m}_{F150LP}-\rm{m}_{F435W}$. Padova isochrones for metallicity $Z=0.004$ and ages between 3 Myr and 10 Gyr are shown in different colors. We assumed the distance modulus DM = 28.02 from \citet{Sabbi2018}, and the two-parameter extinction law formalisim from \citet{Gordon2016}, with $E(B-V)=0.07$, total-to-selective extinction ratio $R(V)=2.7$ and mixing parameter $ f_A=0$.}
\label{f:NGC4449_CMD_ages}
\end{figure*}

{In this paper, we use the nearby \citep[4.01 Mpc;][]{Sabbi2018} irregular barred starburst dwarf galaxy NGC~4449 (morphology SBm) to showcase the GULP data. The galaxy is characterized by a distorted morphology, with ongoing intense star formation (SF) along its central bar (see Fig.~\ref{f:GULPvsGALEX}). Deep H{\sc i} maps reveal an anomalous structure with counter-rotating gas and a halo extending to six times the galaxy's Holmberg radius that could be the result of a recent interaction with another galaxy \citep{Hunter1998} or a minor merger \citep{Lelli2014}. The star formation rate \citep[$\mathrm{SFR}=0.5\, \mathrm{M_\odot\, yr^{-1}}$][]{Lee2009} is more than three times higher than the average value for local galaxies of similar mass \citep[e.g.,][]{Whitaker2012}. The SF is highly concentrated around the central starburst, where the SFR surface density rises to $\Sigma_{\mathrm{SFR}}=0.03\, \mathrm{M_\odot\, yr^{-1}\, kpc^{-2}}$ \citep{Calzetti2018}. 

From HST BVI photometry, several authors concluded that the most recent episode of SF likely occurred between 5 and 10 Myr ago \citep[e.g.,][]{McQuinn2010, McQuinn2012, Sacchi2018}. By taking advantage of HST observations in the U filter F336W, which is more sensitive to the age of the younger stellar populations, \citet{Cignoni2018} inferred a slightly older age for the most recent episode of SF, finding an increase of a factor of $\sim 2$ in the SFR between 10 and 16 Myr ago compared to the average value over the last 100 Myr. Using JWST observations, \citet{Correnti2025} reported for the first time a variation in the density of the 10 and $60\, {\mathrm{Myr}}$ old stars from the North to the South that extends to the clearly distorted Northern spiral arm, suggesting that the extended and most recent episode of SF could be driven by external interactions or tidal effects. Finally, \citet{Meena2025} used the GULP catalog presented in this paper to investigate the hierarchical properties of NGC~4449's young stellar complexes, and their evolution with time.

In the area covered by GULP, NGC~4449  average oxygen abundance is $12 + \log(\mathrm{O/H})\sim 8.2\pm 0.09$ \citep{Berg2012, Pilyugin2015, Annibali2017}, about 30\% of the solar value. This low metallicity is consistent with the low average dust content estimated from the IR/UV ratio, implying that only $\sim 40\%$ of the light from young stars is absorbed by dust \citep{Grasha2013}. Similarly, \citet{Reines2008} and \citet{McQuaid2024} reported $A_V<1.5$ mag, even for embedded sources.} 

Previous studies of NGC~4449 stellar content \citep{McQuinn2010,McQuinn2012,Sacchi2018,Cignoni2018} adopted the typical MW extinction law from \citet{Cardelli1989}. However, when we included the UV filters F150LP and F218W, we found that an extinction law characterized by a weaker ``UV-bump'' at $217.5\, \mathrm{nm}$ and a steeper slope in the FUV better reproduces the colors that include these two filters. 

{We inspected a number of color-magnitude diagrams (CMDs) to interpret the properties of NGC~4449 stellar populations. In making the diagrams we only included sources with photometric error $\rm{MAGER}<0.5$, $-0.4<{\rm SHARP}<0.4$, and detected in more than one }{\tt flt}. {No selection criteria in $\chi^2$ or crowding were used, since they did not appeared to be as effective as the sharpness selection.}

To better reproduce the behavior of the extinction at the shorter wavelengths, we used the python package {\tt dust\_extinction} \citep{Gordon2024} and the two-parameter relationship from \citet{Gordon2016, Gordon2024b}. We adopted a total-to-selective extinction ratio $\rm{R(V)}=2.7$ and mixing parameter $\rm{f_A}=0$, where $\rm{R(V)}$ provides a measurement of the average grain size, and the mixture parameter $\rm{f_A}$ represents the fraction of carbon in the form of Polycyclic Aromatic Hydrocarbons (PAHs) versus graphite, and directly controls the strength of the ``UV-bump'' at $217.5\, \mathrm{nm}$. In the MW diffuse ISM $\rm{f_A}$ is typically one, while in the Small Magellanic Cloud (SMC) is usually zero.

In Figure~\ref{f:extinction_laws} we plot the 3 Myr Padova isochrone \citep{Bressan2012, Chen2014, Chen2015, Tang2014, Marigo2017, Pastorelli2019, Pastorelli2020} for metallicity $\rm{Z}=0.004$, with distance modulus $\rm{DM} = 28.02\, \mathrm{mag}$ and $\rm{E(B-V)}=0.07\, \mathrm{mag}$ \citep[as in ][]{Sabbi2018} over the observed relative star densities in the color-magnitude planes (a.k.a., Hess diagrams) $\rm{m_{F435W}}$ vs. $\rm{m_{F150LP}-m_{F218W}}$, $\rm{m_{F150LP}-m_{F435W}}$, $\rm{m_{F218W}-m_{F435W}}$, and $\rm{m_{F435}-m_{F815W}}$ to highlight how different extinction laws (obtained from different combinations of  $\rm{R(V)}$ and $\rm{f_A}$) behave at different wavelengths. In the upper plots we considered extinction laws characterized by the same slope ($\rm{R(V)}=3.1$) and variable amount of PAH vs. graphite (regulated by the variable $\rm{f_A}$). A ``bumpless'' extinction law (red curves) better reproduces the observations at wavelengths shorter than $\sim 230\ \mathrm{nm}$. In the lower plots, we compared ``bumpless'' extinction laws ($\rm{f_A=0}$) with different slopes (variable $\rm{R(V)}$). In the absence of the ``UV-bump'', we find that the UV colors are better reproduced by a steeper extinction law (i.e., dark red isochrones).

Having constrained the average extinction law, we used Fig.~\ref{f:NGC4449_CMD_ages} to investigate how different color combinations better characterize various stellar populations. The Hess diagrams range from the optical $\rm{m_{F814W}}$ versus $\rm{m_{F555W}-m_{F814W}}$ on the upper-left panel, to the FUV $\rm{m_{F814W}}$ versus $\rm{m_{F150LP}-m_{F435W}}$ on the lower-right. As in Figure~\ref{f:extinction_laws}, we used Padova isochrones for metallicity $\rm{Z}=0.004$, with distance modulus $\rm{DM} = 28.02\, \mathrm{mag}$ and $\rm{E(B-V)}=0.07\, \mathrm{mag}$, but in this plot we used the extinction law determined by $\rm{f_A=0}$, and $\rm{R(V)}=2.7$, and ages between $3\, \mathrm{Myr}$ and $10\, \mathrm{Gyr}$. 

The optical $\rm{m}_{F814W}$ vs. $\rm{m}_{F555W}-\rm{m}_{F814W}$ Hess diagram (upper-left panel), is consistent with that presented by the LEGUS team \citep{Sabbi2018}. It displays a distinct blue plume ($\rm{m}_{F555W}-\rm{m}_{F814W}<0.5$ and $\rm{m}_{F814W}<23.5$), primarily populated by stars in the main sequence and the hot edge of the core helium-burning evolutionary phases, alongside the corresponding red plume ($\rm{m}_{F555W}-\rm{m}_{F814W}>1.5$), consisting of red supergiant (RSG) stars at brighter magnitudes, and asymptotic giant branch (AGB) stars at fainter luminosity.

The well delineated concentration of stars redder than $\rm{m}_{F555W}-\rm{m}_{F814W}\gtrsim 1.0$ and fainter than $\rm{m}_{F814W}>24.05$ is populated by low-mass, old stars in the red giant branch (RGB)
evolutionary phase. The clear drop in stellar counts above  $\rm{m}_{F814W}<24.05$ marks the position of the tip of the red giant branch, caused by the helium flash.  
The red cloud ($2.5<\rm{m}_{F555W}-\rm{m}_{F814W}<4.0$) of stars between $23.8<\rm{m}_{F814W}<24.0\, \mathrm{mag}$ corresponds to the thermally pulsing AGB stars. This Hess diagram covers the longer look-back time, and includes both stars that are just a few Myr old, and sources that are several Gyr years old.

As already highlighted by the LEGUS team \citep{Calzetti2015, Sabbi2018, Cignoni2018}, bluer filters allow to better determine the ages of the younger stellar populations, at the cost of a shorter look-back time. For example, the oldest stars in the $\rm{m}_{F814W}$ vs. $\rm{m}_{F336W}-\rm{m}_{F435W}$ diagram (upper-central plot) are less then $\lesssim 1\, \mathrm{Gyr}$. However, the blue edge of the helium-burning stars is better  separated from the main sequence phase, and looks like a broad cloud of stars redder than $\rm{m}_{F336W}-\rm{m}_{F435W}>0$. The $\rm{m}_{F814W}$ vs. $\rm{m}_{F275W}-\rm{m}_{F435W}$ diagram (upper-right plot) includes only intermediate and massive stars younger than $\lesssim 100\, \mathrm{Myr}$, either in main-sequence or at the blue edge of the helium-burning loop.

The two lower panels in Fig.~\ref{f:NGC4449_CMD_ages} show the diagrams obtained including the two bluest filters: $\rm{m}_{F814W}$ vs. $\rm{m}_{F218W}-\rm{m}_{F435W}$, and $\rm{m}_{F814W}$ vs. $\rm{m}_{F150LP}-\rm{m}_{F435W}$. These diagrams cover a very short look-back time ($\lesssim 30\, \mathrm{Myr}$), but the larger separation between the individual isochrones, and the better sensitivity to the dust extinction, allow a more accurate estimate of stellar ages and masses. 

\subsection{Physical Parameters of the Stars}
\label{sec:SEDs}

\begin{figure*}
\begin{center}
\includegraphics[width=0.7\textwidth]{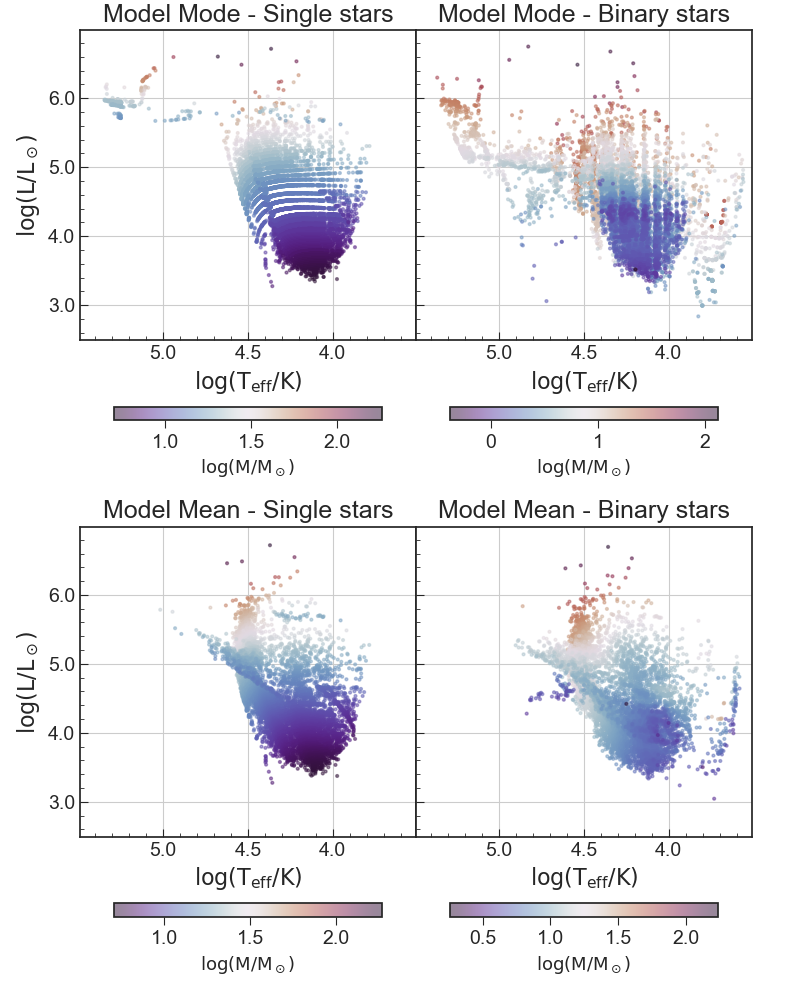}
\caption{HRDs derived from SED fitting of 9034 stars in NGC~4449 with photometric errors $<0.1\, \mathrm{mag}$ in at least 5 broadband filters. Fits were performed with BPASS v2.2, using the same distance modulus, metallicity, and 2-parameter extinction law inferred in Section \ref{sec:CMDs}. {{Upper left:}} Effective temperatures ($\log(\rm{T}_{\rm{eff}}/\rm{K})$) and luminosities ($\log(\rm{L}/\rm{L}_\odot)$) corresponding to the fit PDF mode using single-star evolution only. {{Upper right:}}
Fit PDF mode for effective temperatures and luminosities obtained including binary-star evolution.
{{Lower plots:}} same as above, but showing the fit mean values. Sources are color-coded based on their derived present-day mass. For binary systems, only the primary star is shown.}
\label{f:HRD_1}
\end{center}
\end{figure*}

\begin{figure*}
\begin{center}
\includegraphics[width=0.7\textwidth]{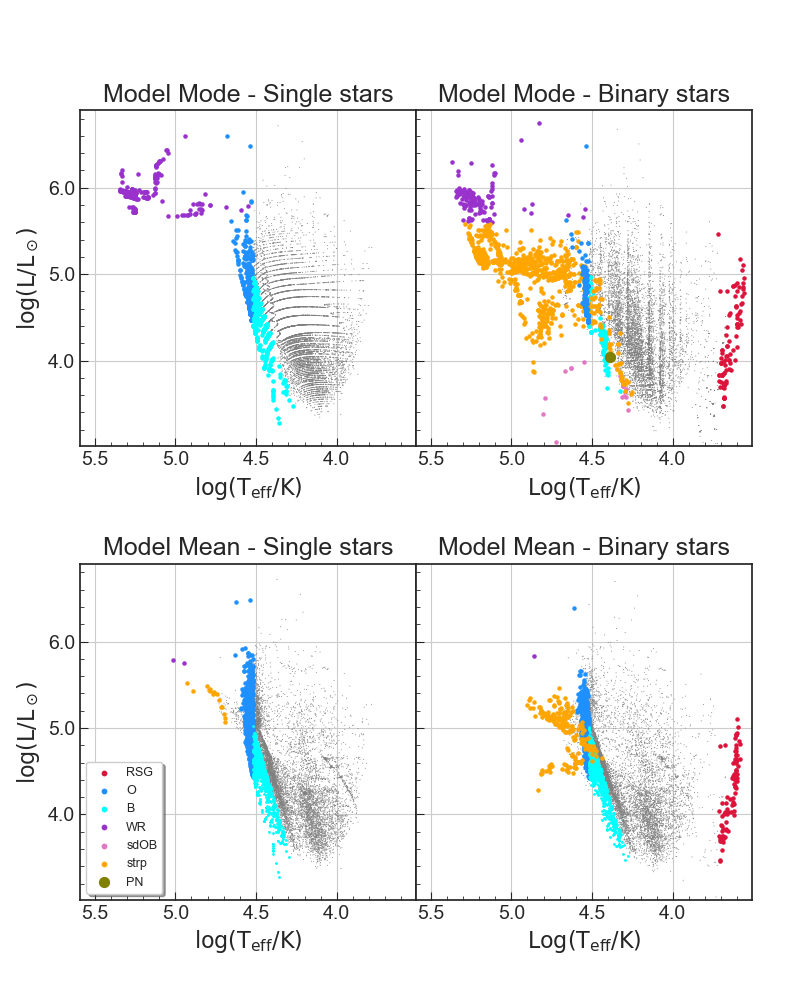}
\caption{The same HRDs shown in Fig.~\ref{f:HRD_1}, with candidate WR stars marked in purple. O-type stars are in blue, B-type in cyan, RSG in red, stripped stars in orange, and sdOB in pink. In the HRD showing the mode of the fit the solution including binary star evolution found a candidate planetary nebula, marked with a large olive dot.}
\label{f:HRD_2}
\end{center}
\end{figure*}

\begin{figure*}
\plotone{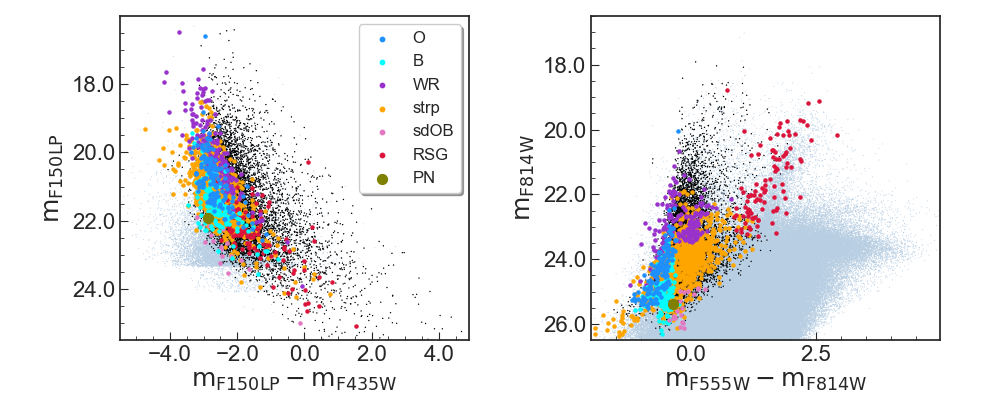}
\caption{Left panel: $\rm{m}_{F150LP}$ vs. $\rm{m}_{F150LP}-\rm{m}_{F435W}$ CMD with superimposed stars in different evolutionary stages as selected from the binary star best-fit solution, shown in the upper right panel of Fig.~\ref{f:HRD_2}. The entire dataset is shown in light gray, and stars for which the SED converged are in black. As in Fig.~\ref{f:HRD_2} candidate O-type stars are in blue, B-type are in cyan, WR in purple, stripped stars are in orange, sdOB in pink, binaries with a RSG companion are in red, and stellar core of the PN in olive. Right panel: same as the left panel but for the $\rm{m}_{F814W}$ vs. $\rm{m}_{F555W}-\rm{m}_{F814W}$ filter combination.}
\label{f:NGC4449_CMD_pops}
\end{figure*}

To derive the masses, temperatures, and ages of the stars found in NGC~4449, we used a custom SED fitting program that leverages on the publicly available Binary Population and Spectral Synthesis code \citep[BPASS v2.2,][]{Stanway2018} and stellar population synthesis models for interacting binaries \citep{Eldridge2017, Stanway2018}. For this analysis, we considered only stars that were  detected in at least five broad-band filters with photometry errors $<0.1\, \mathrm{mag}$. This reduced our catalog to a total of 19153 objects. 

{The BPASS models simulate the evolution of a stellar population originating from a single burst of SF. The models take into account metallicity, the slope of the IMF, and a realistic population of binary stars, based on the initial mass-dependent binary fractions, the initial mass ratio distribution, and the initial period distribution from \citet{Moe2017}. For the IMF in this paper we used a broken power-law with $dN/dM=-1.35$ between 0.1 and $0.5\, M_\odot$, and $dN/dM\propto-2.35$ between 0.5 and $300\, M_\odot$. 

For each burst of SF, BPASS generates an Hertzsprung-Russell diagram (HRD) and the corresponding CMDs. Each HRD consists of a grid, where each cell is populated with the number of stars expected at the given time, based on the assumed IMF. In total, we created 41 HRDs, covering the age-range $6.0\leq \text{log(age/yr)}\leq 10.0$, with a time step of 0.1 dex. The corresponding theoretical CMDs were derived using the same distance, metallicity, and two-parameter $\rm{R(V)},\, \rm{f_A}$ extinction law used in Section \ref{sec:CMDs} (although we ran tests considering also the typical MW extinction) to take into account the impact of dust at the various wavelengths.  

For each star, we identify in each HRD the cell that best matches the observations, and calculated the probability of finding the star in that cell. By combining all the solutions, we derive the probability distribution function (PDF) for each source. Each solution includes the star's initial mass, age, A(V), radius, luminosity, effective temperature, and present-day mass. Furthermore, as a byproduct of the stellar evolutionary models, we also obtain estimates for the masses of the He, CO, and ONe cores, as well as the surface abundances of these elements. 

BPASS simulations can include only single stars, or also binary systems. When the binary evolution is taken into account, BPASS provides the likelihood that a source is a single star or part of a binary (in which case it is classified as merger, star+star binary, star+remnant binary, or unbound binary). When a star is classified as a binary, the fit additionally estimates the initial mass-ratio and orbital period, the present-day orbital period, and the physical parameters of the companion. When considering only single stars, the SED fitting successfully converged for 9034 objects. When we included the binary evolutions the fit converged for 9037 sources.} 

The HRD diagrams obtained using the mode and the mean of the PDF are shown in Fig.~\ref{f:HRD_1}. The results obtained including only single stars are on the left. Those considering also binary evolution are on the right. Uncertainties associated with the various parameters derived by BPASS and correlations about various parameters are shown in Appendix 

As already discussed in Section~\ref{sec:CMDs}, requesting a detection in at least some of the UV filters biases our analysis towards the hotter and younger stars. For the solution considering only single stars, we find that 95\% of sources are hotter than $T_{eff}>10,000\, {\rm K}$, 78\% are still more massive than $8\,{\rm{M}}_\odot$, and 92\% are younger than 50 Myr. 

When we include binary evolution, the number of stars younger than $50\, {\mathrm{Myr}}$ drops to 78\%. In 92\% of the cases the primary star has an effective temperature $T_{eff}>10,000\, {\rm K}$, and 48\% of the companions are also hotter than $T_{eff} > 10,000\, {\rm K}$. Mass transfer appears to play a major role on the mass of the systems: 65\% of the primary stars and 35\% of the companions likely had initial mass above $8\, \rm{M}_\odot$, but only 24\% of the primary stars are still above this mass. On the other hand, the number of companions with present-day mass above $8\, \rm{M}_\odot$ has raised to 45\%.

We used the derived mass, temperature, luminosity, surface gravity,  surface hydrogen (X), and He mass in the core to identify sources in different evolutionary phases (see Fig.~\ref{f:HRD_2}). In particular we defined {O-type star} candidates as all the sources with $\log(\rm{T_{eff}/ K})>4.52$, $\log(\rm{g/cm\,  s^{-2}})>3.0$, and $\rm{X}>70\%$. In the single star best solution 2.6\% of the sources meet these selection criteria. When we allow the binary evolution to be considered, the number drops to 1.5\%. Their position in the HRDs is marked by blue dots.

{B-type star} candidates are defined as the sources with $4.0<\log(\rm{T_{eff}/ K})\leq 4.52$, 
$\log(\rm{g/ cm\, s^{-2}})>3.8$, and $\rm{X}>70\%$. They constitute 4\% of the stars in the single-star best solution and 1.5\% of the sources in the binary one. In Figure~\ref{f:HRD_2} they are marked by cyan dots.

The remaining 88\% to 92\% of the stars are already evolved off the main sequence. In both the single and binary best-fit solutions we find that 2 to 3\% of the sources have SEDs consistent with those of {Wolf-Rayet (WR) stars}. To be classified as a candidate WR, a star must have $\log(\rm{T_{eff}/ K})>4.0$, $\log(\rm{L/ L_\odot})>5.6$, $\log(\rm{g/ cm\, s^{-2}})>3.0$, and $\rm{X}<40\%$. In Figure~\ref{f:HRD_2} WR candidates are marked with purple dots, and represent 2.9\% and 1.8\% of the sample in the single and binary solution respectively. 

In the binary evolution best-fit solution a number of sources are interpreted as the result of interactions and mass-transfer between the two companions. For example, for nearly 10\% of the sources the best-fit solution indicates that ($\log(\rm{T_{eff}/K})>4.0$, $\log(\rm{g/ cm\, s^{-2}})>3.0$, $\log(\rm{L/ L_\odot})<5.6$,  $\rm{M_1}> 1.5\, {\rm M_\odot}$, and $\rm{X}<40\%$, suggesting that the primary star in these  objects lost a major fraction of its hydrogen-envelope. We flagged these systems as {stripped-envelope star} candidates. In Figure~\ref{f:HRD_2} they are marked with orange dots. In 19 cases, the best-fit solution reports a surface hydrogen abundance $\rm{X}<10\%$, which makes these sources candidate {sub-dwarf OB (sdOB)}. In Figure~\ref{f:HRD_2} they are marked with pink dots. 

While single {red supergiants (RSGs)} are likely too cold to be detected in the F336W and bluer filters, for about 100 sources (1.1\% of the sample) the SEDs can be fitted as the combination of a RSG with $\log(\rm{T_{eff}/ K})<3.72$, and either $\rm{X}<10\%$ or mass of the helium core $\rm{M_{HeC}}>0.2\, \rm{M}_\star$, plus a companion with mass between 8 and $32\, \rm{M}_\odot$ and effective temperature between 17,000 and 30,000 K. RSG in binary systems with hot B-type stars have been recently spectroscopically confirmed in the LMC and SMC \citep{Neugent2020, Patrick2022, Patrick2025}. The RSG candidates are marked with red dots.
Finally, in the binary best-fit solution, we find one source with $\log(\rm{T_{eff}/ K})>4.0$, $\log(\rm{g/cm\,s^{-2}})>3.0$, $\rm{X}<0.4$, and $\rm{M_1}< 1.5\, {\rm M_\odot}$ like the sdOB, but $\log(\rm{L/ L_\odot})>4.0$, making it the potential central star of a {planetary nebula (PN)}. This object is marked with a large olive-green dot.

The position of the different types of stars on the FUV CMD $\rm{m_{F150LP}}$ vs. $\mathrm{m_{F150LP}}-\mathrm{m_{F435W}}$ and in the optical CMD $\rm{m_{F814W}}$ vs. $\rm{m_{F555W}}-\rm{m_{F814W}}$ is shown in Fig.~\ref{f:NGC4449_CMD_pops}. The stars are plotted following the same color scheme used in Fig.~\ref{f:HRD_2}. In both CMDs, O-type candidates show very blue colors ($\langle \rm{m}_{\rm{F150LP}}-\rm{m}_{\rm{F435W}}\rangle_O=-2.66$, and $\langle \rm{m}_{\rm{F555W}}-\rm{m}_{\rm{F814W}}\rangle_O=-0.64$ respectively). This is also true for the B-type candidates ($\langle \rm{m}_{\rm{F150LP}}-\rm{m}_{\rm{F435W}}\rangle_B=-2.55$ and $\langle \rm{m}_{\rm{F555W}}-\rm{m}_{\rm{F814W}}\rangle_B=-0.56$). 

The WR candidates are among the bluest sources in the FUV CMD ($\langle \rm{m}_{\rm{F150LP}}-\rm{m_{F435W}}\rangle_{WR}=-2.65$) and span a range of more than 6 magnitudes in the F150LP filter ($17.6<\rm{m_{F150LP}}<23.9$), with an average magnitude of $\langle \rm{m_{F150LP}}\rangle_{WR}=20.02$. In the optical, the WR candidates can be divided into two groups: a bluer one with an average color of $\langle \rm{m_{F555W}-m_{F814W}}\rangle_{WR}=-0.04$, and a redder one with $\langle \rm{m_{F555W}-m_{F814W}}\rangle_{WR}=0.07$. For both groups the F814W luminosity varies between $21.4 < \rm{m_{F814W}}<24.8$, with an average value of $\langle \rm{m_{F814W}}\rangle_{WR}=22.93$. For comparison, the O-type candidates' mean magnitudes is $\langle \rm{m_{F150LP}}\rangle_O=20.792$ in the FUV, and $\langle \rm{m_{F814W}}\rangle_O=24.38$ in the optical.

Candidate stripped stars have colors similar to those of the WR ($\langle \rm{m_{F150LP}-m_{F435W}}\rangle_{strp}=-2.75$ and $\langle \rm{m_{F555W}-m_{F814W}}\rangle_{strp}=-0.29$, respectively), but are considerably fainter ($\langle 
\rm{m_{F150LP}}\rangle_{strp}=21.19$, and $\langle \rm{m_{F814W}}\rangle_{strp}=24.24$). Finally, the candidate RSGs are among the faintest and redder sources in the UV ($\langle \rm{m_{F150LP}}\rangle_{RSG}=22.58$, and $\langle \rm{m_{F150LP}-m_{F435W}}\rangle_{RSG}=-1.27$), while in the visible they are among the brightest ($\langle \rm{m_{F814}}\rangle_{RSG}=21.65$) and reddest ($\langle \rm{m_{F555W}-m_{F814W}}\rangle_{RSG}=1.52$) sources. 

\section{NGC~4449's Star Clusters}
\label{sec:YSCs}

\begin{figure*}
    \centering
    \includegraphics[width=0.9\linewidth, height=0.68\textheight, keepaspectratio]{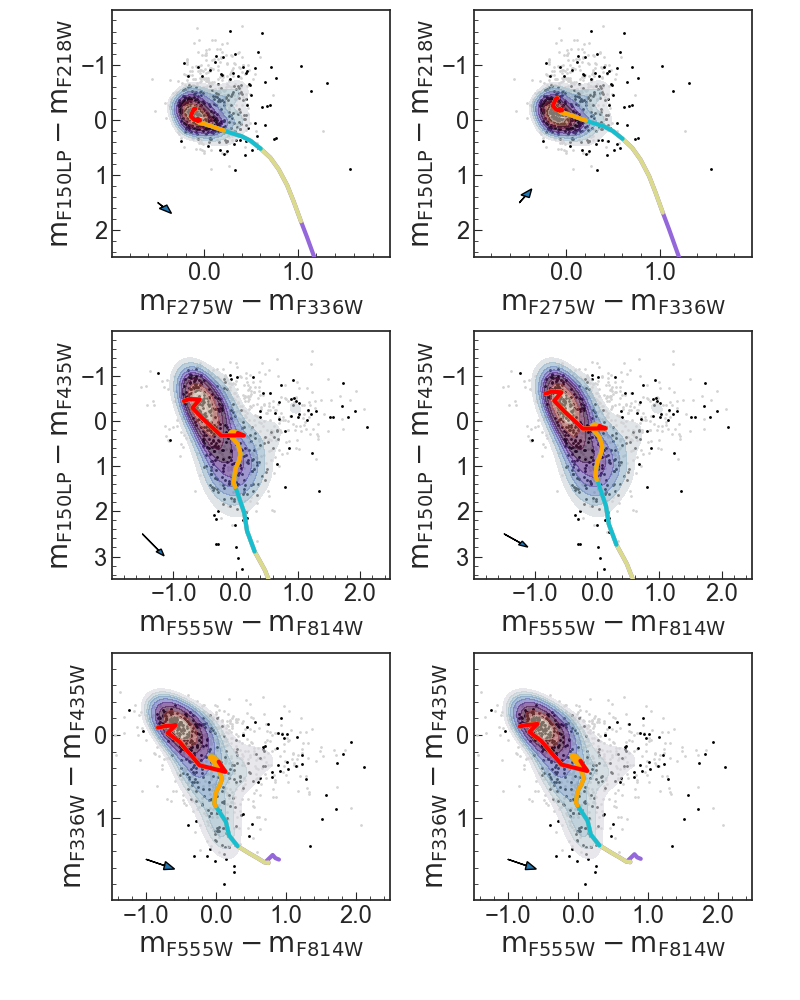}
    \caption{\small Color-color diagrams for the 1224 automatically extracted sources in the LEGUS catalogue of NGC~4449, detected in the GULP filters F150LP and F218W with photometric error $<0.2\, {\mathrm{mag}}$, after foreground extinction removal. Black dots mark the 396 Class~1, 2, and 3 YSC candidates; Class~4 interlopers are shown in gray. Overlaid KDEs represent regions where the YSC probability density is $>20\%$ of the maximum. Padova evolutionary tracks ($Z=0.008$, Kroupa IMF, nebular covering factor $C_f=0.5$) are shown for LMC (left) and MW (right) extinction laws assuming $E(B-V)=0.08$~mag. Color-coding for ages: red ($<10\, {\rm Myr}$), orange (10--$100\, {\rm Myr}$), cyan (100--$500\, {\rm Myr}$), olive ($500\, {\rm Myr}$--$1\, {\rm Gyr}$), and purple ($>1\, {\rm Gyr}$).}
    \label{f:clusters_CC}
\end{figure*}
\begin{figure*}
    \centering
    \includegraphics[width=1.0\linewidth]{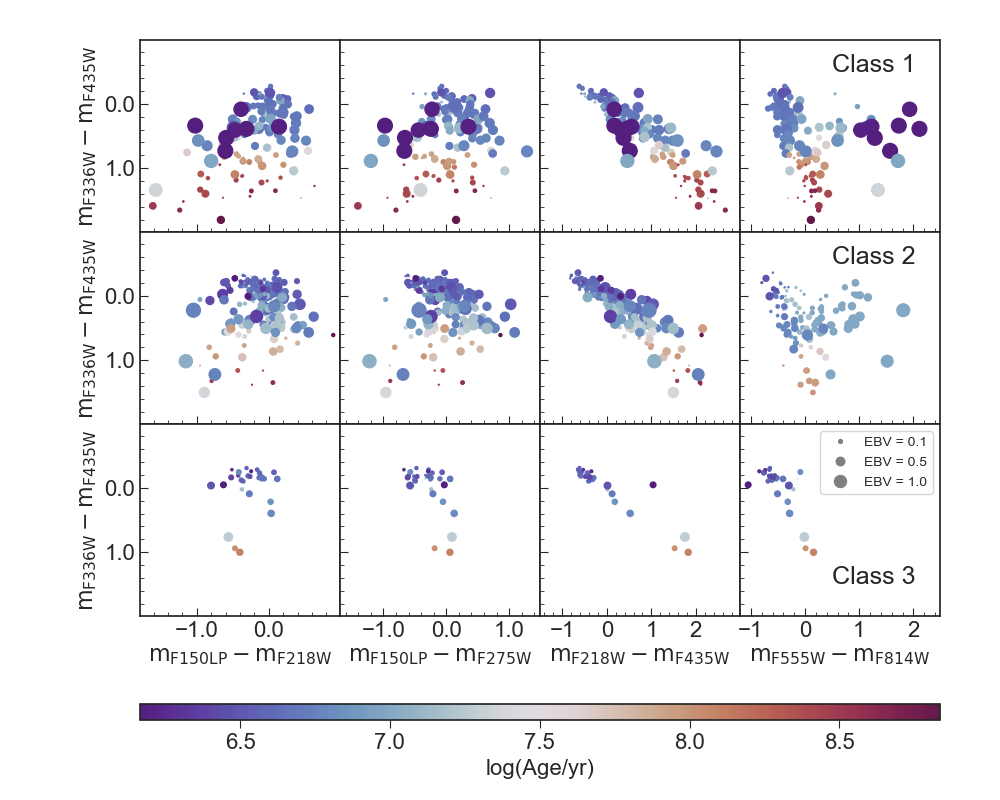}    \caption{Color-color diagrams for 369 sources classified as Class~1, Class~2, or Class~3 clusters in NGC~4449. The upper row shows the 127 sources classified as Class 1, the 206 Class 2 sources are shown in the middle row, and the remaining 30 Class~3 clusters are in the lower plots. Sources are color-coded by age, with blue being younger and red older. The size of the markers is proportional to the $\rm{E(B-V)}$ value, with smaller markers are affected by lower extinction.}
    \label{f:cc_classes}
\end{figure*}
The NGC~4449 LEGUS catalog contains 2853 sources \citep{Cook2019, Cook2023}. Of these, 1734 were detected also in the filters F150LP and F218W, 1222 with photometric errors $<0.2\, {\mathrm{mag}}$. The different size between the two catalogs is the result of both the smaller area covered by the F150LP filter, and the shorter look-back time. In fact, the LEGUS sample includes clusters as old as 14 Gyr, while the oldest cluster in GULP is only 2 Gyr. Of the 1222 sources in the GULP catalog, 369 were visually classified in the LEGUS catalog as Class~1, 2, or 3 star cluster candidates. We will focus our analysis only on these objects.

Figure~\ref{f:clusters_CC} shows examples of the clusters' color-color diagrams for different combinations of filters, after the correction for foreground extinction \citep[$\rm{A_V}=0.052\, {\mathrm{mag}}$, ][]{Schlafly2011}. The kernel density estimate (KDE) contours outline the regions where the probability density of Class 1, 2, and 3 cluster candidates is higher than 20\%. A Padova evolutionary track for metallicity Z=0.008 and color-coded for different age ranges is shown for reference. In the plots on the left side of the figure, we used an LMC attenuation curve with $\rm{E(B-V)}=0.08\, {\rm mag}$, while for those on the right we used the MW attenuation curve. As in the case of the field stars, the FUV colors are better reproduced by an attenuation curve with a weaker ``UV-bump'' and a steeper rise in the FUV.  

{In order to derive the star clusters physical properties, we fitted the SED of each target using the publicly available Code Investigating GALaxy Emission \citep[CIGALE, ][]{Boquien2019} package, a forward-modeling code that operates on a grid of theoretical models spanning different star formation histories, stellar populations, and dust attenuation curves. Stellar mass is constrained through the overall normalization of the stellar population models to the observed fluxes, and a Bayesian statistical analysis is used to calculate the PDF. Final parameter estimates are taken as the likelihood-weighted mean over the model grid, with uncertainties derived from the width of the corresponding PDFs.

For this analysis, we used the \citep{Bruzual2003} stellar library with a double-exponential star formation history and a \citet{Chabrier2003} initial mass function. Extinction curves for the MW, LMC, and SMC were adopted from \citep{Pei1992}, with the amplitude of the ``UV-bump'' at 217.5 nm ranging from 0 to 5 (where 3 corresponds to the MW). To better constrain the ages of the younger clusters, we included photometry from the H$\alpha$ F658N filter in the fitting process. 

Figure~\ref{f:cc_classes} shows a series of color-color diagrams, for the individual classes of cluster candidates. In these cases, the YSCs are color-coded based on their age, as derived from the CIGALE SED fitting,} with younger clusters being blue and older ones red. The size of the markers is proportional to the amount of intrinsic reddening E(B-V). 

The Class~1 clusters (four upper panels) represent 34\% of the entire GULP sample and typically exhibit redder and more dispersed colors (i.e., $\langle \rm{m_{F150LP}-m_{F275W}}\rangle = 0.01\pm 0.46$ and $\langle \rm{m_{F555W}-m_{F814W}}\rangle = 0.00\pm 0.57$). This is likely due to a combination of larger amount of internal reddening ($\langle \rm{E(B-V)}\rangle=0.27\pm 0.34\, \rm{mag}$) and older ages ($\langle \rm{Age}\rangle=71 \pm 239\, \rm{Myr}$). The Class~1 cluster candidates are also the most massive systems in the sample, with $\langle \text{M}\rangle = (0.6 \pm 3)\times 10^5\, \mathrm{M_\odot}$. 

The Class~2 clusters, shown in the central four panels, have intermediate colors (i.e., $\langle \rm{m_{F150LP}-m_{F275W}}\rangle = -0.08\pm 0.37$ and $\langle \rm{m_{F555W}-m_{F814W}}\rangle = -0.2\pm 0.5$) and on average appear to be affected by a lower amount of intrinsic extinction ($\langle \rm{E(B-V)}\rangle = 0.13\pm 0.21\, {\mathrm{mag}}$). The Class~2 clusters represent 56\% of the GULP sample, have intermediate ages ($\langle \text{age}\rangle = 45 \pm 543\, {\mathrm{Myr}}$) and masses ($\langle \text{M}\rangle = (1 \pm 4)\times 10^4\, \mathrm{M_\odot}$).

Only a few ($\sim 8\%$) Class~3 clusters are detected in the F150LP and F218W filters. These systems, identified also as OB associations in \citet{Adamo2017}, are typically smaller ($\langle \text{M}\rangle = (3 \pm 1)\times 10^3\, \mathrm{Myr}$) and younger ($\langle \text{age}\rangle = 1 \pm 14\, {\mathrm{Myr}}$) than the rest of the sample, and exhibit relatively blue colors ($\langle \rm{m_{F150LP}-m_{F275W}}\rangle = -0.29\pm 0.22$, $\langle \rm{m_{F555W}-m_{F814W}}\rangle = -0.49\pm 0.27$), with small  intrinsic reddening ($\langle \rm{E(B-V)}\rangle = 0.08\pm 0.18\, {\mathrm{mag}}$).

\section{Spatial and Temporal Progression of Star Formation in NGC~4449}
\label{sec:SFH}

\begin{figure*}
\plotone{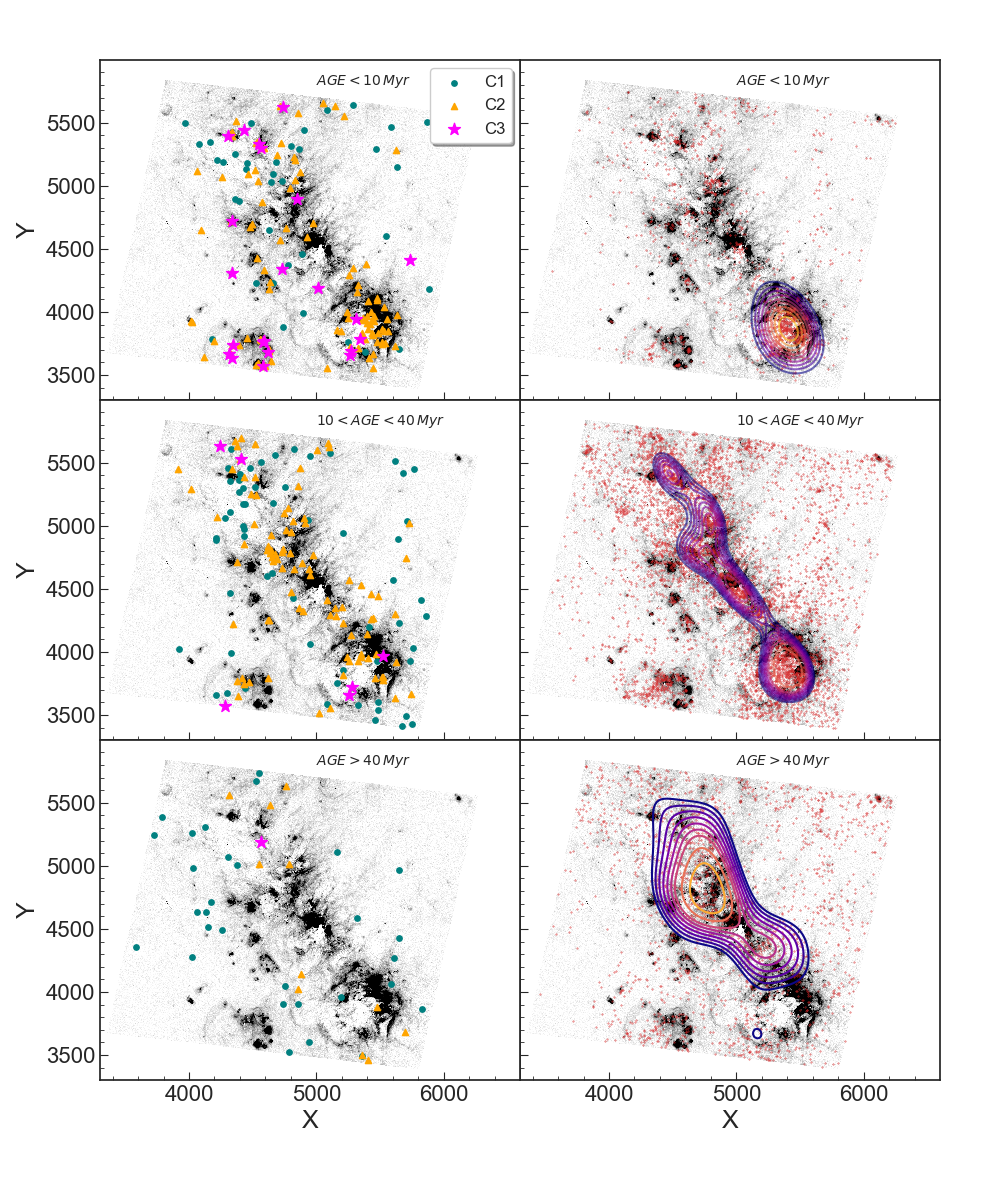}
\caption{Spatial distribution of clusters (left panels) and stars (right panels) for different age ranges overlaid upon the continuum subtracted portion of the F657N image that is covered also by the F150LP observations. On the left Class~1 objects are marked with green dots, Class~2 with orange triangles, and Class~3 with magenta stars. On the right, individual massive stars are shown as red dots, while KDE contours delineate the regions of the galaxy where the probability density of sources is higher than 68\% of the peak density.}
\label{f:age_dist}
\end{figure*}

YSCs and massive stars can easily be identified  even in distant galaxies, because of their significantly higher luminosity than older populations. Furthermore, because of their young ages, most of the massive stars have not had enough time to migrate far away from their formation site, and therefore, they effectively map the spatial and temporal occurrence of SF across a galaxy. This is shown, for example, in Figure~\ref{f:age_dist} where the distribution of stars and clusters in different age bins is over-plotted on the NGC~4449 continuum-subtracted F658N image, {which was created by subtracting a synthetic stellar continuum (derived via linear interpolation of the F555W and F814W broadband fluxes) from the F658N data}.

The majority of the stars younger than $\lesssim 100\, \mathrm{Myr}$ are predominantly found along NGC~4449 bar, with the locus of SF migrating from the North-East towards the South-West, in agreement with recent results presented by \citet{Correnti2025}.
The two upper panels show that during the past $\sim 10\, {\mathrm{Myr}}$, most of the SF occurred within a large ($D\sim 25\arcsec$, corresponding to 484 pc at 4.01 Mpc) cavity toward the South-West (x=5500, y=3850; RA=12:28:09, Dec=44:05:11). The high concentration of Class~2 clusters (orange triangles) younger than $10\, {\mathrm{Myr}}$ is likely responsible for blowing away most of the ionized gas that shines as a bright ring in the F658N image.

The majority of the YSCs and stars between 10 and $40\, {\mathrm{Myr}}$ (central panels) appear to be tightly associated with the central bar-like structure, while stars and clusters older than $40\, {\mathrm{Myr}}$ (lower panels) exhibit a wider distribution, though still aligned with the bar. We ascribe this change in density to the expansion of the star-forming regions under the random stellar motion and dispersion with a possibly greater cluster dispersal rate inside the bar than outside. If the bar also had a higher star formation rate density than the surrounding disk, then this explanation would account for the higher fraction of young clusters in the bar and the relatively high fraction of old clusters outside the bar. A detailed analysis of the hierarchical properties of the young stellar complexes and their evolution with time is discussed in \citet{Meena2025}. 

The relatively rapid dispersion of stellar complexes in the bar of the galaxy is further confirmed by the distribution and properties of the older clusters, shown in the lower-left panel. Almost all the clusters older than $> 40\, {\mathrm{Myr}}$ belong to the Class~1 type (teal dots), which consists of the most massive, gravitationally bound clusters. The older cluster candidates are found predominantly in the disk of the galaxy, outside the galaxy bar, while the, typically younger, Class~2 and 3 systems are found mostly close to the bar and other regions of high ionization. A similar distribution with Classes~3 and 2 still closely associated with the ionized gas, and the Class~1 found farther away from the most recent sites of SF was also reported for M51 by \citet{Grasha2019}, although in that case clusters should move into and out of the spiral arms more easily than clusters in NGC~4449 can move into and out of the bar. Even in the older age group, stars remain more concentrated in the bar than in the rest of the galaxy disk, compared to clusters of similar age. This further support the hypothesis that clusters are more rapidly destroyed in the bar than outside \citep[e.g., ][]{Kruijssen2011, Berentzen2012, Rossi2015}. This seems to be the best evidence that clusters in the bar are preferentially getting destroyed (and turning into field stars still in the bar) compared to clusters outside the bar, which are still present despite being are old.

\section{Dust extinction and attenuation in NGC~4449}
\label{sec:UV-bump}

\begin{figure*}
\plotone{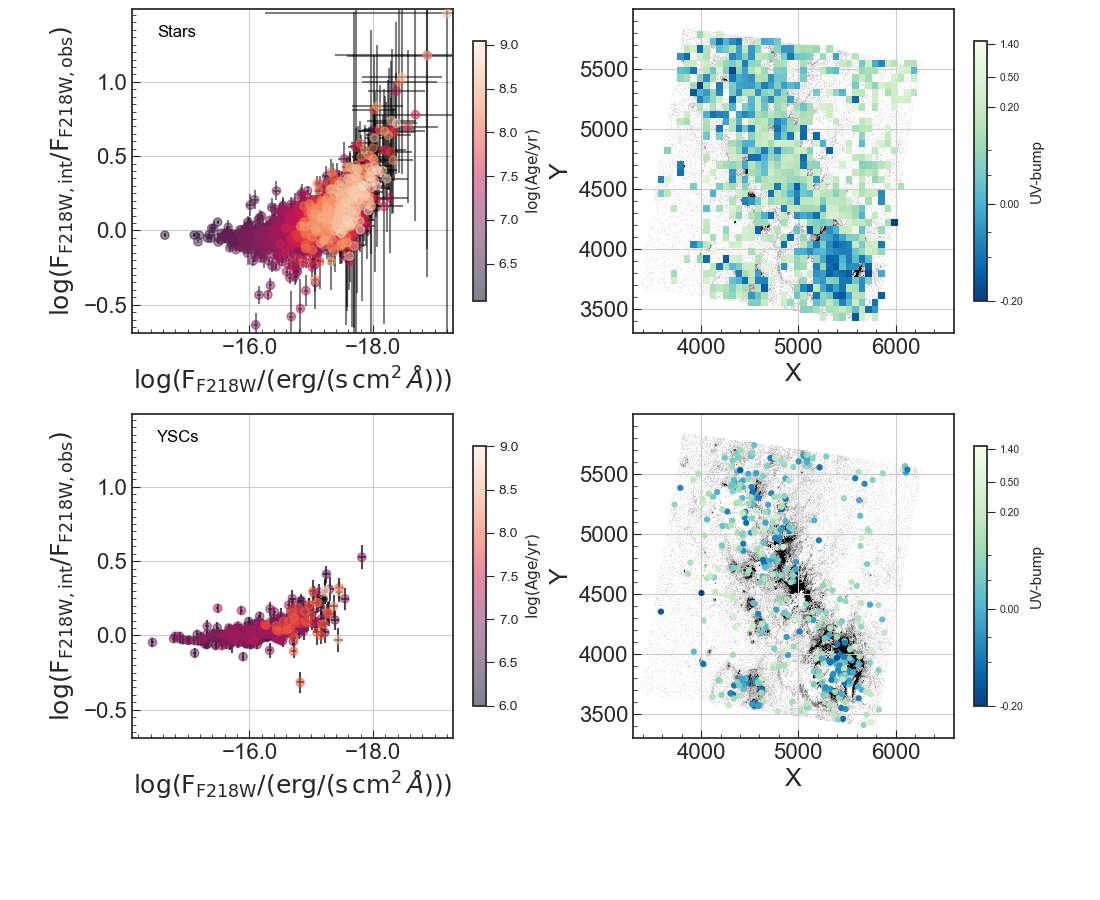}
\caption{``UV-bump'' strength and spatial variation for stars and clusters in NGC~4449. The upper-left panel shows the ratio between the predicted and observed star flux in the filter F218W (i.e., strength of the ``UV-bump'' at $\lambda=2175\, \mathrm{\text{\AA}}$), with the predicted flux derived by interpolating the stellar fluxes  from the filters F150LP and F275W as a function of the flux in the F218W filter. Sources are color-coded based on their age, which was estimated from the binary-solution SED fitting. The upper-right panel shows the spatial variation of the median strength  of the ``UV-bump'' across the galaxy, as derived from stellar photometry, displayed with an arcsine hyperbolic normalization. The lower-left panel shows the strength of the ``UV-bump'' for the YSC candidates, while the lower-right panel shows distribution of the YSCs across NGC~4449. As for the stars, the YSCs are color-coded based on the strength of the ``UV-bump'', using an arcsine hyperbolic normalization.}
\label{f:NGC4449_UVBump}
\end{figure*}

Dust is a minor component in mass of the ISM. Nevertheless, interstellar grains have a major role in the evolution of galaxies, the formation of stars and planets, and the synthesis of complex organic molecules, possibly associated with the origin of life \citep{Draine2011, Klessen2016}

UV extinction curves have been determined along several lines of sight in the MW, as well as in the MCs and in M31. In most cases \citep{Valencic2004}, the MW extinction curve can be well described by a relationship that depends only on the parameter R(V)=A(V)/E(B-V) \citep[i.e.]{Cardelli1989, Mathis1992, Fitzpatrick2019}. However, in the metal-poor MCs, the attenuation curve in YSCs often exhibits a steeper UV slope and a smaller ``UV-bump'' at $\lambda=2175$ \AA\, \citep{Prevot1984, Gordon1998, Misselt1999, Gordon2003, Maiz-Apellaniz2012, Gordon2016, Pang2016, DeMarchi2016, Gordon2024b}. ``Bumpless'' attenuation curves are often found also in starburst environments and Lyman break galaxies, both at low and high redshift \citep{Calzetti1994, Calzetti2000, Calzetti2005, Vijh2003, Scoville2015, Salim2019}.

The strength of the `UV-bump'' and the steepness of the attenuation curve in the UV are attributed to the chemical composition, geometry, and size of the dust grains. The origin of the absorption feature at $\lambda=2175$ \AA\, is still not well understood \citep{Draine1989, Draine2003}, but nowadays is often attributed to nano-particles containing PAH \citep{Joblin1992, Bradley2005, Shivaei2022}, or graphite \citep{Li2001}. A small ``UV-bump'' combined with a steep attenuation curve in the UV is attributed to a high abundance of small dust grains that favor higher scatter efficiency.  

Extinction curves with variable UV steepness and ``UV-bump'' strength have been used to characterize the distribution, size, coating composition of dust grains, and the presence of PAH molecules at $0<z<3$ \citep[e.g.,][]{Noll2007, Noll2009, Shivaei2022}. However,  the UV-steep/``bumpless'' extinction curves towards Trumpler~37 \citep{Valencic2003}, and other deviations towards dark clouds and bright nebulae in the MW \citep{Mathis1992} highlight a more complex scenario for dust formation and disruption. Furthermore, the inclination of a galaxy can severely affect the interpretation of its dust properties. For example, the almost edge-on starburst galaxy M82 shows a clear `UV-bump'' \citep{Hutton2014, Brown2015}, likely due to the intervening disk of the galaxy. This effect can be further exacerbated at high-redshift, due to the limited spatial resolution.

Here we show how HST observations in the filters F150LP, F218W, and F275W can be used to create star-to-star maps of the ``UV-bump'' strength for thousands of sources and hundreds of YSCs, and we discuss how these parameters vary as a function of local conditions. To derive the strength of the ``UV-bump'', we first corrected the photometry for foreground extinction \citep[assuming $\rm{A(V)} = 0.052\, {\mathrm{mag}}$ from the ][ maps]{Schlafly2011}. We then defined the intensity of the ``UV-bump'' $\log(\mathrm{F}_{\mathrm{F218W, int}}/\mathrm{F}_{\mathrm{F218W, obs}})$, as the logarithmic difference between the expected and the observed flux of a source in the filter F218W. For the expected flux $\mathrm{F}_{\mathrm{F218W, int}}$, we used the interpolation between the fluxes  in the filters F150LP and F275W.

Dolphot produces photometric catalogs in the Vegamag system. To convert the stars' magnitudes into fluxes (in units of $\mathrm{erg/(s\, cm^{2}\, {\text{\AA}})}$) we used the PHOTFLAM values reported in the image headers.

We measured the photometry of the YSCs directly in the ABmag system. To convert their magnitudes into $\mathrm{erg/(s\, cm^{2}\, \text{\AA})}$, we used the standard equation $\log(F_\lambda) = -0.4(m_\lambda + 48.6) + \log(2.9979 \times 10^{18} / PHOTPLAM_\lambda^2)$ is the filter pivot wavelength.

Figure~\ref{f:NGC4449_UVBump} shows how the strength of the ``UV-bump'', $\log(\mathrm{F}_{\mathrm{F218W, int}}/\mathrm{F}_{\mathrm{F218W, obs}})$, varies with the UV-flux for stars (upper panels) and YSCs (lower panels). For both types of objects the strength of the ``UV-bump'' remains negligible for sources brighter than $\log(\mathrm{F_{F218W}/(erg/(s\, cm^{2}\, \text{AA}})))<-16$. Below this value, the ``UV-bump'' is almost absent in stars close to intense UV-emitters, but becomes progressively stronger as the distance from the UV-emitters increases. This is consistent with the hypothesis that intense UV-radiation destroys the PAH dust grains responsible for the ``UV-bump'' \citep[e.g., ][]{Gordon2008, Wu2011}. The anti-correlation between the spatial distribution of stars and clusters with a strong ``UV-bump'' and regions of intense H$\alpha$ emission (plots on the right side of Fig.~\ref{f:NGC4449_UVBump}) further highlights the impact of the radiation field on the chemistry of the ISM. A discussion on the effect of SF activity and metallicity on the evolution of dust and ISM in general is presented in \citet{Galliano2018},  

\section{Summary}
\label{sec:conclusions}

GULP is an HST Cycle 28 Treasury program that imaged 26 nearby star-forming galaxies in the far and near UV using the ACS/SBC filter F150LP and the WFC3/UVIS filter F218W. When combined with archival observations, GULP provides a panchromatic 8-band view of these systems from the FUV to the I band. The galaxies were selected to cover a broad range of metallicities, masses, morphological types, and star formation rates. By providing a view of regions of massive star formation in nearby galaxies with unprecedented high-spatial resolution in the FUV and NUV, GULP significantly expands our understanding of massive stars, OB associations, and YSCs in nearby star-forming galaxies. 

GULP provides a rich dataset for unraveling the intricate interplay between massive stars, YSCs, and the ISM in diverse galactic environments. GULP will be particularly important for quantifying the impact of intense UV radiation on dust properties and the evolution of dust grains in different environments, which is essential for interpreting observations of galaxies at higher redshifts.

Our analysis of NGC 4449 shows that young stellar populations ($\lesssim 100\, \mathrm{Myr}$ old) are predominantly concentrated along the galaxy's central bar, showing a migration of star formation activity from the northeast to the southwest on a timescale of a few tens of Myrs. Similarly, younger clusters are tightly associated with regions of recent star formation, whereas older clusters and field stars exhibit a wider distribution, suggesting that stellar systems in the disk and in the bar are subject to different destruction rates, as suggested by N-body simulations for the evolution of star clusters in barred galaxies \citep[e.g., ][]{Rossi2015}. A more in depth study of the clustering properties of NGC~4449 massive stars is presented in \citep{Meena2025}, while in \citet{Facchini2025}, we analyze the properties of candidate isolated O-stars, and discuss their possible origin.

The inclusion of the F150LP and F218W filters allowed us to directly measure and map the strength of ``UV-bump''$\log(\mathrm{F}_{\mathrm{F218W, int}}/\mathrm{F}_{\mathrm{F218W, obs}})$ at $\lambda=2175\,\mathrm{\text{\AA}}$. We show that the ``UV-bump'''s spatial distribution mirrors the regions of most intense ionization, where it is weakest. This finding strongly supports the hypothesis that the intense UV radiation within young, active star-forming regions effectively photodissociates small dust grains, leading to a weaker or absent ``UV-bump''. In a forthcoming paper we will extend this investigation to metal-poor dwarf galaxies characterized by more modest UV radiation, to better understand the roles of metallicity and UV radiation in regulating ISM properties.

\appendix
\label{app}

\begin{figure*}
\begin{center}
\includegraphics[width=1\textwidth]{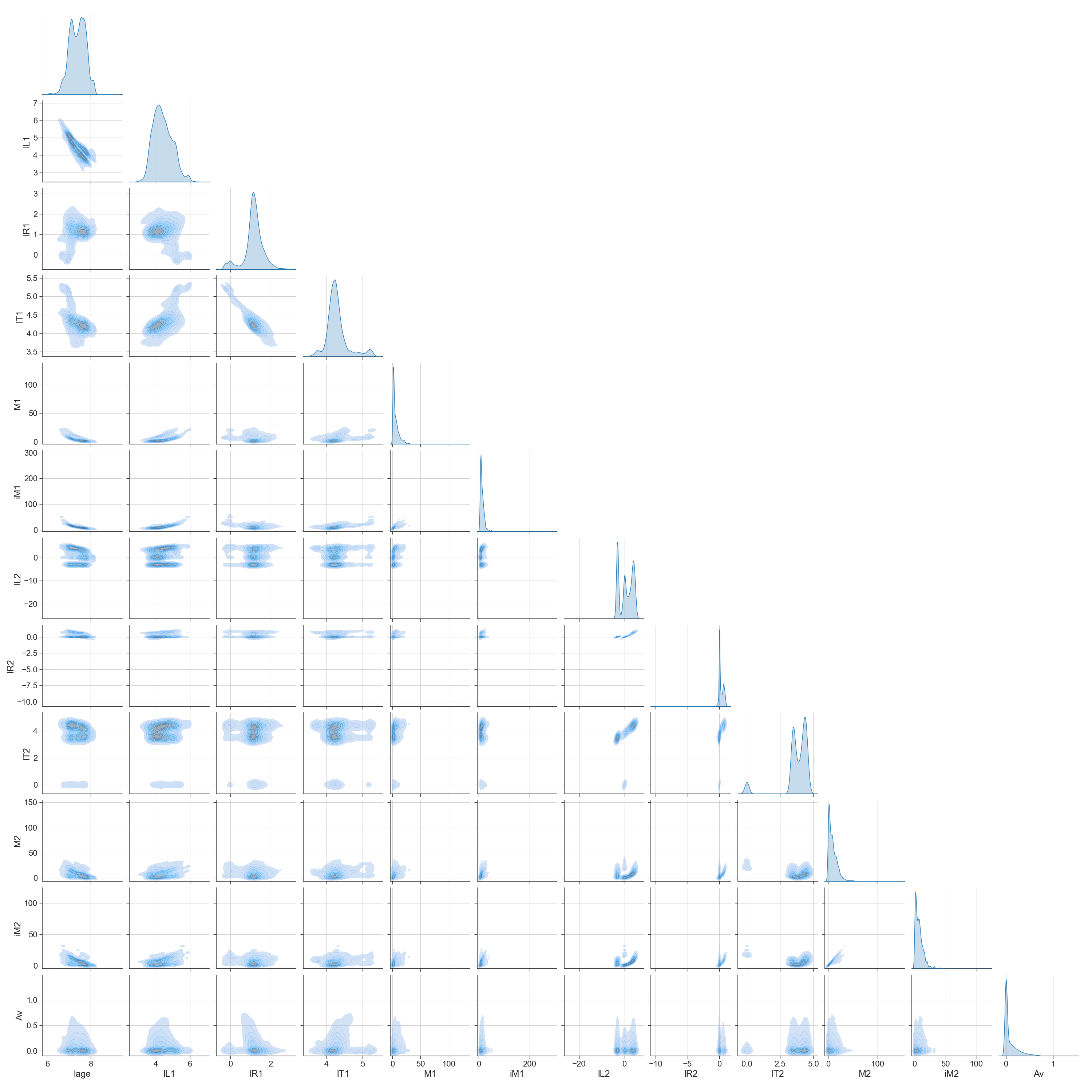}
\caption{Correlations between the mode of the PDF for each source fitted with BPASS. From left to right, and from top to bottom the plot shows: the $\text{log}(\text{Age/yr})$ (lage) of the source, the luminosity $\text{log}(\text{L}_1/\text{L}_\odot)$ (lL1), the radius $\text{log}(\text{R}_1/\text{R}_\odot)$ (lR1), the effective temperature $\text{log}(\text{T}_1/\text{T}_{\text{eff}})$ (lT1) in logarithmic scale, the present day mass (M$_1$) and the initial mass (iM$_1$) of the primary star, the luminosity $\text{log}(\text{L}_2/\text{L}_\odot)$ (lL2), the radius $\text{log}(\text{R}_2/\text{R}_\odot)$ (lR2), the effective temperature $\text{log}(\text{T}_2/\text{T}_{\text{eff}}$ (lT2) in logarithmic scale, the present day mass (M$_2$) and the initial mass (iM$_2$) of the secondary star, and A(V).
}
\label{f:BPASS_mode_param}
\end{center}
\end{figure*}

\begin{figure*}
\begin{center}
\includegraphics[width=1\textwidth]{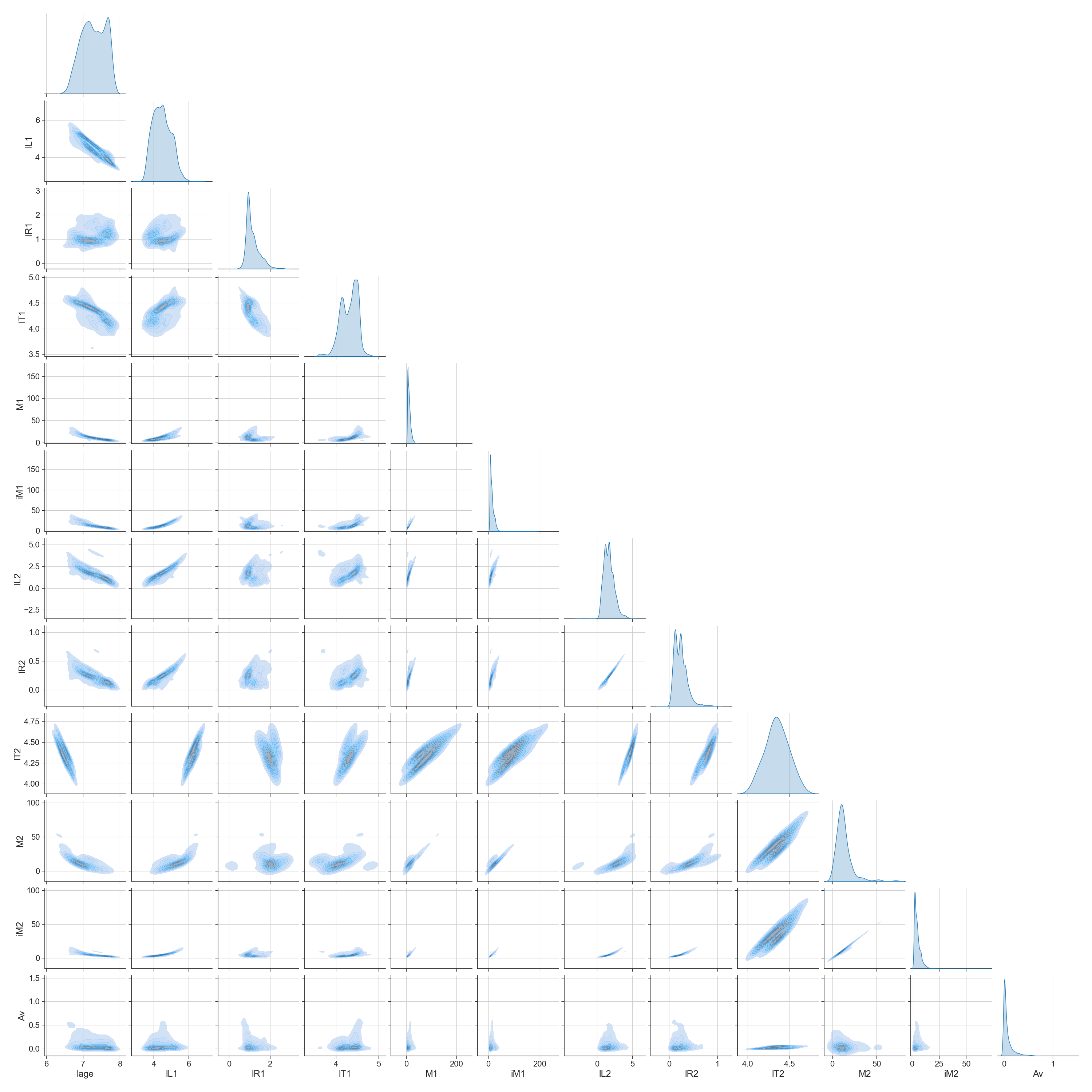}
\caption{The same as Figure~\ref{f:BPASS_mean_param}, but for the PDF mean values.
}
\label{f:BPASS_mean_param}
\end{center}
\end{figure*}

\begin{figure*}
\begin{center}
\includegraphics[width=1\textwidth]{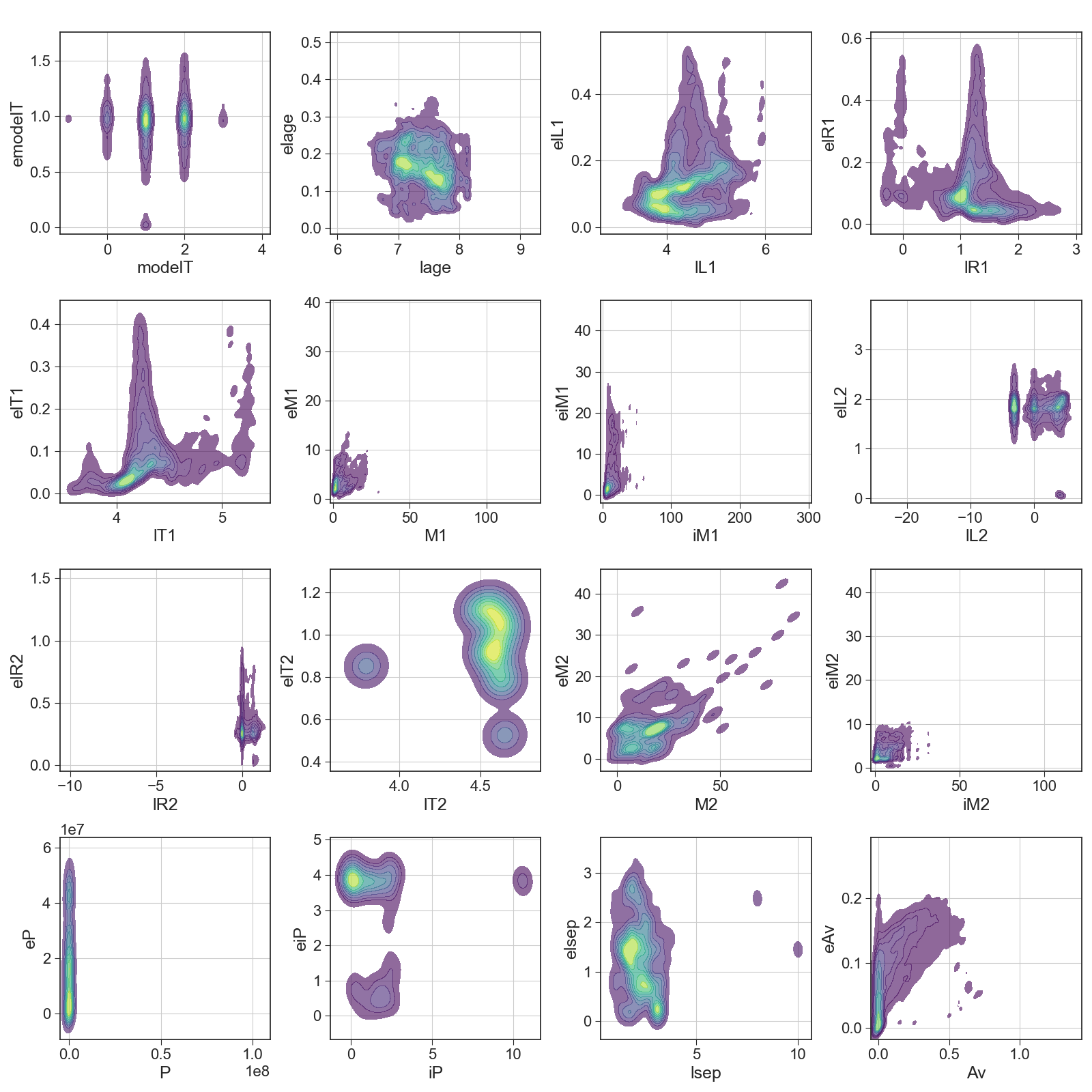}
\caption{PDF mode uncertainties for the model type (-1 = single star, 0 = binary merger, 1 = binary system, 2 = binary system with one compact object, 3 = unbound companion, 4 = quasi homogenelus evolution objects), age, luminosity, radius, effective temperature, present mass and initial mass for the primary and secondary stars, present period (P), inital period (iP), separation ($\log(\text{sep})$) and A(V).   }
\label{f:BPASS_mode_error}
\end{center}
\end{figure*}

\begin{figure*}
\begin{center}
\includegraphics[width=1\textwidth]{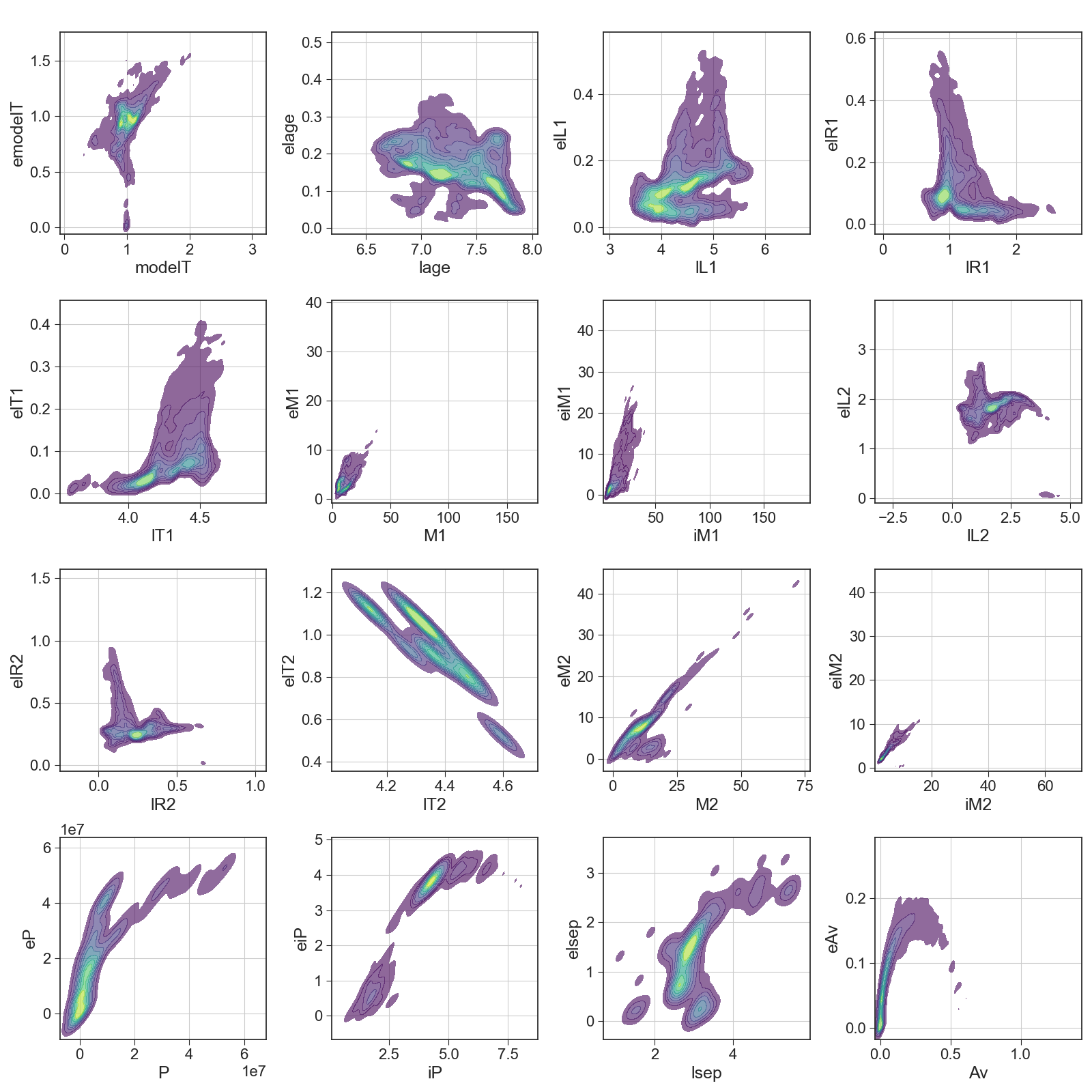}
\caption{The same as Figure~\ref{f:BPASS_mode_error}, but for the PDF mean.   }
\label{f:BPASS_mean_error}
\end{center}
\end{figure*}

Figure~\ref{f:BPASS_mode_param} shows the correlations among the PDF mode values of various parameters provided by BPASS, similarly  the PDF mean values for the same parameters are shown in Fig.~\ref{f:BPASS_mean_param}. The uncertainties of PDF mode and mean value obtained for various parameters are derived by BPASS are shown in Figs.~\ref{f:BPASS_mode_error} and  \ref{f:BPASS_mean_error}, respectively.

\setlength{\hoffset}{-0.5in} 
\movetabledown=1.7in
\begin{rotatetable*}
\startlongtable
\begin{deluxetable*}{llcccccccccc}
\setlength{\tabcolsep}{0.02in}
\tabletypesize{\footnotesize}
\captionsetup{width=0.4\textwidth}\tablecaption{Target list. The first column reports the target name, the morphological type is in column 2, column 3 reports the UV SFR, the galaxy mass in in column 4, metallicity is in column 5, and the distance in column 6. Column 7 reports the WFC3 filters (eihter than F218W) that were acquired by GULP because not already available in the archive, and column 8 the archival ACS filters that were used when WFC3 filters were not available.}
\tablehead{
\colhead{Name}  & \colhead{Morph} & \colhead{T} & \colhead{SFR(UV)}            & \colhead{M$_\star$}   & \colhead{12+log(O/H)} & \colhead{Dist}  & \multicolumn{3}{c}{GULP} & \multicolumn{2}{c}{Archive}\vspace{-2ex} \\ 
\colhead{} &    \colhead{} &      \colhead{} & \colhead{($M_\odot yr^{-1}$)} & \colhead{(M$_\odot$)} & \colhead{}            & \colhead{(Mpc)} & \colhead{SBC pnt} & \colhead{WFC3 filters} &  \colhead{WFC3 pnt} & \colhead{WFC3}  & \colhead{ACS}
} 
\startdata
IC~4247  & S?                            & 2.2 & 0.008 & 1.2E8 & 8.27$^a$ & 5.11 & 8 &  F657N & 1 & & F606W, F814W \\
NGC~4258 & \multirow{2}{*}{SABbc} & \multirow{2}{*}{4.0} & \multirow{2}{*}{2.51}  & \multirow{2}{*}{2.9E10} & \multirow{2}{*}{8.89$^a$} & \multirow{2}{*}{6.83} & \multirow{2}{*}{8} &       & \multirow{2}{*}{1} & & \\    
(M106)     &  &  &  &  &  &  &  &        &  & F657N & \\
NGC~4605 & SBc   & 5.1 & 0.457 & 1.38E9 & 8.77$^a$ & 5.56 & 8 &        & 1 &  F657N    &        \\
NGC~5457 & \multirow{2}{*}{SABcd} & \multirow{2}{*}{6.0} & \multirow{2}{*}{6.71} & \multirow{2}{*}{1.9E10} & \multirow{2}{*}{8.48$^c$} & \multirow{2}{*}{6.7} & \multirow{2}{*}{18} &  & \multirow{2}{*}{2} & & \\
(M101)  &  &  &  &  &  &  &  &        &  &  & \shortstack{F435W, F555W \\ F658N, F814W}\\
NGC~1313 & SBd   & 7.0 & 1.15  & 2.57E9 & 8.4$^a$  & 4.3 & 18 &  & 2 & F657N & F435W, F555W, F814W\\
NGC~4242 & SABdm  & 7.9 & 0.1   & 1.07E9 & 7.5$^{\star}$ & 5.3 & 9 &  & 1 & F657N &  \\
NGC~5474 & SAcd  & 6.1 & 0.27  & 8.13E8 &  8.31$^b$          & 6.6 & 9 &  & 1 & F657N & F606W, F814W\\
NGC~7793 & SAd   & 7.4 & 1.92  & 3.24E9 &  8.31$^b$          & 3.8 & 16 &  & 2 & F657N & F435W, F555W, F814W \\
UGC~685  & SAm   & 9.2 & 0.007 & 9.55E7 & 8.00$^c$           & 4.37 & 8 & F657N & 1 & & F606W, F814W \\
NGC~4395 & SAm   & 8.9 & 0.339 & 6.03E8 & 8.26$^b$           & 4.52 & 16 &  & 1 & F657N & F555W, F814W\\
UGC~4305 & Im    & 9.9 & 0.12  & 2.29E8 & 7.92$^c$           & 3.32 &  8 &  & 1 & & F550M, F658N, F814W \\
UGC~4459 & Im    & 9.9 & 0.007 & 6.67E6 & 7.82$^c$           & 3.96 & 8 &  F657N & 1 & & \\
UGC~5139 & IABm  & 9.9 & 0.012 & 2.5E7  & 8.00$^c$           & 3.83 & 8 &  F657N & 1 & & F555W, F814W \\
NGC~3738 & Im    & 9.9 & 0.07  & 2.4E8  & 8.04$^c$           & 5.09 & 9 &  & 1 & & F606W, F658N, F814W\\
MRK~209  & \multirow{2}{*}{Sm} & \multirow{2}{*}{10} & \multirow{2}{*}{0.02} & \multirow{2}{*}{1.9E7} & \multirow{2}{*}{7.82$^c$} & \multirow{2}{*}{5.19} &  \multirow{2}{*}{4} &  & \multirow{2}{*}{1} & F657N & F606W, F814W\\
(UGC~A281)   &  &  &  &  &  &  &  &        &  & &  \\
NGC~4449 & IBm   & 9.8 & 0.94  & 1.1E9 & 8.26$^c$  & 4.01 & 9 &  & 1 & & \shortstack{F435W, F555W \\ F658N, F814W}\\
NGC~5253 & Im    & 11  & 0.1   & 2.19E8 & 8.25$^c$ & 3.32 & 4 &  & 1 & F657N & F435W, F555W, F814W \\
\multirow{2}{*}{NGC~300}  & \multirow{2}{*}{Sc}    & \multirow{2}{*}{6.9} & \multirow{2}{*}{0.32}  & \multirow{2}{*}{1.9E9}  & \multirow{2}{*}{8.57$^d$}  & \multirow{2}{*}{1.88$^*$} & \multirow{2}{*}{-} & \multirow{2}{*}{F657N} & \multirow{2}{*}{1} & \multirow{2}{*}{F218W, F225W} & F150LP, \\
&  &  &  &  &  &  &  &        &  & & F435W, F555W, F814W \\
UGC~A292 & \multirow{2}{*}{dSPh} & \multirow{2}{*}{9.9} &  \multirow{2}{*}{0.002} & \multirow{2}{*}{3.98E5} & \multirow{2}{*}{7.30$^e$} & \multirow{2}{*}{3.1$^*$} & \multirow{2}{*}{4} &  \multirow{2}{*}{F657N} & \multirow{2}{*}{1} & & \multirow{2}{*}{F475W, F606W, F814W}\\
(CVN-I-DWA)  &  &  &  &  &  &  &  &        &  & &  \\
MCG+0920131 & \multirow{2}{*}{nb/d} & \multirow{2}{*}{10} & \multirow{2}{*}{0.002} & \multirow{2}{*}{6.03E5} & \multirow{2}{*}{6.9$^{\star}$} & \multirow{2}{*}{3.23$^*$} & 1 &  F275W, F438W & 1 & & \\
(CGCG269-049) &  &  &  &  &  &  &  &        &  & &  \\
UGC~9240 & Im & 10 & 0.006 & 1.86E7 & 7.95$^e$ & 2.8 & 8 &  F657N & 1 & & F475W, F606W, F814W\\    
NGC~3741 & Im & 9.9 & 0.068 & 6.4E6 & 7.62$^e$ & 3.19 & 4 & F675N, F555W & 1 & & F475W, F814W \\
UGC~8091 & Im & 9.8 & 0.003 & 1.58E7 & 7.65$^e$ & 2.13 & 4 &  F675N, F555W & 1 &  & F475W, F814W \\
SagDIG & IB(s)m & & & & & 1.08$^{**}$ & 4 &  F675N & 1 & & F475W, F606W, F814W\\
UGC~7242 & Scd & 6.4 & 0.007 & 7.76E7 & 7.32$^{\star}$ & 5.67& 4 &  F657N & 1 &  & F606W, F814W\\
NGC~4485 & IBm & 9.5 & 0.25 & 3.7E8 & 6.92$^{\star}$ & 8.7 & 8 &  & 1 & F657N & F435W, F606W\\
\enddata
    \label{t:targets}
\end{deluxetable*}
\end{rotatetable*}  
\setlength{\hoffset}{0.in} 

\begin{deluxetable*}{ccccccccc}
\tabletypesize{\footnotesize}
\tablecolumns{9}
\tablewidth{0pt}
\tablecaption{ Cross-correlation Fit Details \label{t:results}}
\tablehead{
Ev. Phase & \colhead{ $\log(\rm{T_{eff}/K})$} & 
\colhead{$\log(\rm{g/cm\,s^{-2}})$} & \colhead{$\log(\rm{L/L_\odot})$} & 
\colhead{$H_{s\, ab}$} & \colhead{$iM_1\, [M_\odot]$} & 
\colhead{$M_1\, [M_\odot]$} & \colhead{$iM_2\, [M_\odot]$} & \colhead{$M_2\, [M_\odot]$}}
\startdata
\multicolumn{9}{c}{{Only Single Stars}} \\
\hline
{WR} & $>4.0$ & $>3$ & $>5.6$ & $<40\%$ & 36 -- 300 & 15 -- 92 & -- & -- \\ 
{O-type} & $>4.52$ & $>3.0$ & & $>70\%$ & 17 -- 200 & 17 --183 & -- & -- \\
{B-type} & 4 -- 4.52 & $>3.8$ & & $>70\%$ & 6.3 -- 21 & 6.28 -- 20.8 & -- & -- \\
\hline
\multicolumn{9}{c}{{With Binary Stars}} \\
\hline
{WR} & $>4.0$ & $>3$ & $>5.6$ & $<40\%$ & 19 -- 300 & 12 -- 98 & 1 -- 270 & 1 -- 314 \\ 
{O-type} & $>4.52$ & $>3.0$ & & $>70\%$ & 9.5 -- 150 & 17 -- 133 & 1 -- 32 & 1 --  37 \\
{B-type} & 4 -- 4.52 & $>3.8$ & & $>70\%$ & 6.3 -- 21 & 6.28 -- 20.8 & -- & -- \\
{St-O} & $>4$ & $>3$ & $<5.6$ & $<40\%$ & & $>1.5$ & & \\
{HWD} & $>4$ & $>3$ & $>4$ & $<40\%$ & & $<1.5$ & & \\ 
{sdOB} & $>4$ & $>3$ & $<4$ & $<40\%$ & & $<1.5$ & & \\ 
{RSG} & $<3/72$ &  &  &  & &  & & \\ 
\enddata
\vspace{-0.8cm}
\end{deluxetable*}

\begin{acknowledgments}
We are grateful to the anonymous referee for the careful review and useful suggestions, that helped improving this paper. 

The HST observations used in this paper are associated with program No. 16316. Support for program No. 16316 was provided by NASA through a grant from the Space Telescope Science Institute.

This work is based on observations obtained with the NASA/ESA Hubble Space Telescope, at the Space Telescope Science Institute, which is operated by the Association of Universities for Research in Astronomy, Inc., under NASA contract NAS 5-26555. This work has made use of data from the European Space Agency (ESA) mission {\it Gaia} (\url{https://www.cosmos.esa.int/gaia}), processed by the {\it Gaia} Data Processing and Analysis Consortium (DPAC, \url{https://www.cosmos.esa.int/web/gaia/dpac/consortium}). Funding for the DPAC has been provided by national institutions, in particular, the institutions participating in the {\it Gaia} Multilateral Agreement.

E.S. is supported by the international Gemini Observatory, a program of NSF NOIRLab, which is managed by the Association of Universities for Research in Astronomy (AURA) under a cooperative agreement with the U.S. National Science Foundation, on behalf of the Gemini partnership of Argentina, Brazil, Canada, Chile, the Republic of Korea, and the United States of America.
A.A acknowledges support from the Swedish National Space Agency (SNSA) through the grant 2021- 00108.
R.S.K.\ acknowledges financial support from the ERC via Synergy Grant ``ECOGAL'' (project ID 855130),  from the German Excellence Strategy via the Heidelberg Cluster ``STRUCTURES'' (EXC 2181 - 390900948), and from the German Ministry for Economic Affairs and Climate Action in project ``MAINN'' (funding ID 50OO2206).  R.S.K.\ also thanks the 2024/25 Class of Harvard Radcliffe Fellows for highly interesting and stimulating discussions. 
\end{acknowledgments}

\vspace{5mm}
\facilities{HST(ACS/SBC); HST(WFC3/UVIS); HST(ACS/WFC)}
\software{drizzlepac \url{https://drizzlepac.readthedocs.io/en/latest/index.html} ;
astropy \citep{astropy:2013, astropy:2018}  
           }

\clearpage
\bibliography{GULP}{}
\bibliographystyle{aasjournal}

\end{document}